\documentclass[preprint,12pt,authoryear]{elsarticle}

\usepackage{amssymb}
\usepackage{amsmath}
\usepackage{graphicx}
\usepackage[strings]{underscore}
\usepackage{xurl, hyperref}
\usepackage{listings}
\usepackage{xcolor,verbatim}
\usepackage{tikz}
\usetikzlibrary{
  arrows.meta,     
  positioning,     
  shapes.geometric 
}
\usetikzlibrary{calc} 

\journal{}

\begin{document}

\begin{frontmatter}



\title{Exploratory Modelling of Multi-System Transformation Pathways from Real-World Data:\\ A SINDy-Inspired Sparse Orthogonal Regression Technique}


\author[ijs]{Sabin Roman}
\author[ait]{Vitaliy Soloviy}
\author[ait]{Klaus Kubeczko}

\affiliation[ijs]{
  organization={Jožef Stefan Institute, Department of Knowledge Technologies},
  addressline={Jamova cesta 39},
  city={Ljubljana},
  postcode={1000},
  country={Slovenia}
}

\affiliation[ait]{
  organization={AIT Austrian Institute of Technology GmbH},
  addressline={Giefinggasse 4},
  city={Vienna},
  postcode={1210},
  country={Austria}
}
\begin{abstract}

Sustainability transitions unfold through interacting dynamics across social, technological, economic, environmental, and governance dimensions. However, many modelling approaches either isolate subsystems or rely on optimisation-based pathways that do not explicitly represent feedbacks, path dependence, and institutional constraints. This study develops a Sparse Orthogonal Regression Technique (SORT), a data-driven dynamical-systems framework for reconstructing multi-system transformation dynamics from harmonised European indicators. Inspired by Sparse Identification of Nonlinear Dynamical Systems (SINDy), SORT infers parsimonious cross-domain dependencies from observed data rather than specifying a full structural model ex ante.

The prototype covers energy, emissions, innovation, digitalisation, resource productivity, environmental stress, policy, finance, wellbeing, and resilience. The compact dynamical system is interpreted as a structural representation of medium-term interactions, not as a forecasting tool. It reproduces differentiated empirical patterns, including steady renewable expansion, nonlinear scaling in transition finance, diffusion-like digitalisation, oscillatory environmental stress, and sustained reduction in ETS-regulated emissions. Forward simulations diverge from simple linear extrapolation where reinforcing feedbacks are present, illustrating the conditional nature of accelerated decarbonisation. By translating inferred dependencies into a feedback structure, the analysis identifies reinforcing decarbonisation sequences alongside balancing ecological constraints. The model contributes to sustainability transitions research by providing an empirically grounded representation of multi-system feedback dynamics that bridges quantitative modelling and transition theory. SORT offers a compact technique for exploratory reconstruction from real-world transition data.
\end{abstract}







\end{frontmatter}


\tableofcontents

\section{Introduction}

Accelerating socio-ecological transitions requires the coordination of change across multiple interdependent systems. Energy infrastructure, industrial production, digitalisation, financial flows, governance frameworks, and social well-being evolve through feedbacks that are rarely linear and seldom synchronised. Traditional transition modelling frameworks struggle to capture these interactions with sufficient transparency, contextual fidelity, or temporal relevance \citep{kohler2018modelling,holtz2015prospects}. This report presents preliminary work and an initial prototype of a compact, systematically calibrated, dynamically structured modelling approach that supports the analysis and governance of multi-system transformation pathways in the European context. The approach, empirical grounding, and analytical scope are intended to be further refined and extended through subsequent iterations. The goal is not to forecast deterministic futures, but to isolate general structural tendencies, generate transparent counterfactuals, and provide actionable insight for science–policy interfaces.

The work is positioned within ongoing efforts to develop modelling approaches that integrate industrial, energy and socio ecological dimensions, and that operate across micro, meso and macro scales \citep{truffer2022perspective, andersen2024multi}. The project background emphasises the need to advance conceptual and methodological capacities to model multi system transformations, with particular attention to the linkages between industrial transformation, energy, digitalisation, and the long term sustainability of provisioning systems. In line with this mandate, the prototype focuses on structural and dynamical relationships rather than detailed techno economic optimisation. This approach reflects the observation in transition literature that governance, social legitimacy, system inertia, and non linear responses often dominate pathway feasibility more than marginal cost calculations \citep{geels2022causality}.

A further driver of this research is the need for modelling approaches that reflect multiple relevant temporal scales while remaining empirically grounded. Long term transitions do not evolve smoothly: they progress through periods of acceleration, stalling, crisis and restructuring \citep{rosenbloom2017pathways}. The empirical window used in this study spans 2005 to 2025, and the model is applied to generate stylised projections for an additional 20 years. This horizon up to 2045 is long enough for structural tendencies to manifest, yet still constrained by the available data. Capturing such dynamics requires models that represent qualitative shifts including regime transitions, tipping behaviour, structural bottlenecks and rebound effects using minimal but empirically meaningful parametrisations.

To anchor the qualitative conceptual framework in observed system behaviour, the modelling approach draws on EU wide indicators spanning energy, emissions lock in, resource productivity, digitalisation, innovation, governance, finance, social well being, environmental stress and resilience. These were selected during discussions as core variables reflecting provision system transformations and structural conditions relevant to NEFI pathways. The empirical analyses presented show that distinct variables exhibit distinct dynamical signatures: some evolve almost linearly (for example governance stringency), some exhibit convex acceleration (transition finance), some oscillate (environmental stress), some follow diffusion type growth (digitalisation), and some demonstrate non linear regime shifts (ETS emissions). These empirical dynamics motivate the chosen modelling architecture: a light weight coupled dynamical system that reflects structural causal relationships while respecting observed temporal profiles. The results provide a more realistic representation of transition behaviour than either simple trend extrapolation or rigid optimisation models.

Crucially, the modelling effort is not framed as a forecasting exercise but as a science policy support tool \citep{van2016checklist}. The background explicitly situates the work as providing actionable insights that can inform transformation oriented science policy interfaces. Hence, the outcomes emphasise robustness of qualitative conclusions rather than numerical precision. Such an approach aligns with broader methodological shifts in transition modelling, where model salience and transparency for stakeholders are increasingly valued alongside technical complexity. The prototype therefore emphasises interpretability, modularity, and traceability of causal structure. Its minimality is not a limitation but a feature: it makes the model auditable, teachable, and adaptable for participatory use.

A final motivation concerns feasibility and the need for methodological innovation. The work is framed as a feasibility study to explore a proof of concept that can guide longer term development. This scope justifies focusing on establishing structural validity over exhaustive data parametrisation. The immediate contributions of this work are therefore threefold: (1) extract robust empirical regularities and structural relationships governing multi system transitions; (2) encode these in a compact dynamical formalism that reflects empirically observed temporal characteristics over the period 2005 to 2045; and (3) demonstrate how such a model provides interpretive traction relevant for transformation oriented governance and foresight.

In summary, this report introduces and analyses a prototype dynamical model for multi system transformation pathways. The model is built in explicit alignment with the stated aims of the collaboration, capturing structural relationships across provisioning systems, spanning micro, meso and macro dynamics, reflecting observed temporal behaviour, and delivering interpretable insight for science policy processes. While the model remains intentionally lean, the results reveal clear empirical patterns, illustrate non trivial system interactions, and provide a foundation for developing more elaborated versions in subsequent work. The approach offers a tractable yet expressive alternative to conventional integrated assessment and optimisation models, with the specific advantage of reflecting realistic dynamical structure over policy relevant timescales.

\section{Literature review}

This study is situated at the intersection of sustainability transitions research,
formal modelling of socio-technical change, and emerging work on multi-system
transformations. Rather than seeking predictive precision, the aim is to develop
a structurally informed representation of interdependencies across social,
technological, financial, environmental, and governance domains. This orientation
aligns with a growing recognition that sustainability transitions are neither
single-sector processes nor reducible to isolated technological substitutions,
but instead unfold through coupled dynamics across multiple systems and scales.

The literature reviewed below provides the conceptual and methodological grounding
for this approach. It is organised into four strands: (i) modelling sustainability
transitions and socio-technical change; (ii) causal reasoning and dynamical feedbacks
in sustainability science; (iii) multi-system and provisioning system perspectives;
and (iv) implications for modelling practice and policy relevance.

\subsection{Modelling sustainability transitions}

Formal modelling has become an increasingly important complement to qualitative
and historical approaches in sustainability transitions research. Early transition
scholarship was dominated by narrative case studies and conceptual frameworks,
most notably the multi-level perspective (MLP). While these approaches provided
rich interpretive insights, they faced limitations in systematically exploring
dynamic interactions, counterfactuals, and non-linear feedbacks.

Several reviews and position papers have argued that modelling can address these
limitations by making assumptions explicit and enabling structured experimentation
\citep{kohler2018modelling,holtz2015prospects}. Rather than replacing qualitative
analysis, modelling is increasingly framed as a complementary tool that supports
the exploration of interdependencies, timing effects, and emergent behaviour in
complex transition processes.

System dynamics (SD) has been particularly influential in this regard. SD models
emphasise endogenous change, feedback loops, and stock–flow structures, which
align well with core transition concepts such as lock-in, path dependence, and
regime destabilisation. Early SD applications explicitly linked modelling to
transition theory, demonstrating how regime shifts can emerge from interactions
between incumbent systems, niche innovations, and external pressures
\citep{papachristos2011system}. Subsequent work has highlighted SD’s ability to
explore policy leverage points, actor inertia, and governance dynamics without
assuming optimal or frictionless behaviour \citep{holtz2015prospects}.

More broadly, modelling sustainability transitions has been framed as an
exploratory rather than predictive exercise. \cite{kohler2018modelling}
stress that transition models should be evaluated in terms of their capacity to
represent qualitative system change, co-evolution, and normative tensions, rather
than their forecasting accuracy. This perspective directly informs the modelling
strategy adopted in this study.

\subsection{Causal reasoning and feedbacks in sustainability science}

A central challenge in modelling transitions lies in how causal claims are framed
and justified. Sustainability problems are characterised by intertwined social
and ecological processes, long time horizons, and limited opportunities for
controlled experimentation. As a result, causal inference often relies on a
combination of empirical analysis, theoretical reasoning, and modelling.

Recent contributions have explicitly problematised causal reasoning in this
context. Geels \citep{geels2022causality} argues that socio-technical transitions
require approaches to causality that go beyond simple variable–outcome relations,
emphasising conjunctural causation, event sequences, and feedback-driven dynamics.
Similarly, Schl\"uter et al.\ \citep{schluter2024navigating} highlight the diversity
of causal assumptions across disciplines and the need for transparency when
making causal claims in sustainability research.

From a modelling perspective, this has motivated calls for tighter integration
between empirical analysis and process-based models. Schl\"uter et al.\
\citep{schluter2023unraveling} show how modelling can help interpret empirical
patterns by revealing underlying mechanisms and dynamic constraints, rather
than serving as a substitute for empirical inference. Exploratory modelling
approaches further emphasise systematic variation of assumptions to assess the
robustness of qualitative insights under uncertainty \citep{moallemi2020exploratory}.

These arguments support the use of sparse, dynamically structured models that
focus on dominant dependencies and feedbacks, while avoiding over-interpretation
of higher-order terms or coefficients. In this sense, causal diagrams derived
from regression results should be interpreted as structural summaries rather
than exhaustive causal maps.

\subsection{Multi-system dynamics and provisioning systems}

An important limitation of much transition modelling is its focus on single
systems, such as electricity or transport, treated largely in isolation.
Recent work has increasingly emphasised that sustainability transitions unfold
through interactions between multiple socio-technical systems, including energy,
mobility, industry, finance, and welfare provision.

Andersen and Geels \citep{andersen2023multi} identify a set of causal processes
through which multi-system interactions can accelerate or constrain net-zero
transitions, highlighting the roles of institutional alignment, actor strategies,
and technological complementarities. This work underscores that interactions
between systems are not merely contextual, but can fundamentally shape transition
pathways.

Parallel to this, provisioning system perspectives have gained prominence as a
way to integrate social outcomes, material stocks, and resource flows within a
single analytical frame. Plank et al.\ \citep{plank2021doing} conceptualise
provisioning systems as configurations of infrastructures, institutions, and
practices that jointly deliver societal services. From this perspective, variables
such as wellbeing, productivity, and resilience are not external outcomes, but
integral components of system dynamics.

This orientation resonates with broader debates on transformations to sustainability,
which stress the need to combine structural, systemic, and enabling approaches
\citep{scoones2020transformations}. Rather than privileging technological change
alone, these approaches emphasise the co-evolution of social conditions, governance,
and material systems.

In the context of this study, the inclusion of variables such as social wellbeing,
environmental stress, and resilience reflects this multi-system perspective.
Their interactions with innovation, finance, and policy are treated as endogenous
features of the transition process, not as external shocks.

\subsection{Implications for modelling practice and policy relevance}

A recurring concern in the literature is the relationship between modelling,
policy relevance, and legitimacy. Models that obscure uncertainty or present
overly deterministic narratives risk undermining trust and narrowing political
debate. Stirling \citep{stirling2023mind} explicitly warns against technocratic
uses of models that conceal ignorance and value judgements under the guise of
objectivity.

In response, several authors argue for greater reflexivity and humility in
modelling practice. This includes transparent communication of assumptions,
attention to plural pathways, and recognition of models as partial representations
of complex realities \citep{kohler2018modelling,holtz2015prospects}. From this
viewpoint, causal loop diagrams and system representations should be understood
as heuristic devices that support dialogue and learning, rather than definitive
accounts of how systems function.

The modelling approach adopted in this study follows these principles. By focusing
on dominant dependencies revealed through sparse regression and by curating causal
structures that are both empirically grounded and conceptually coherent, the aim
is to provide a form of ``causal scaffolding'' that can inform scenario development
and policy discussion without claiming predictive authority.

Overall, the literature supports a modelling strategy that is exploratory,
multi-system oriented, and attentive to feedbacks between social, environmental,
and economic domains. This study contributes to this emerging body of work by
demonstrating how data-driven dynamical modelling can be integrated with transition
theory to yield structurally meaningful insights into sustainability transitions.

\section{Methodology}

\subsection*{Data-driven identification of dynamical relationships}

The empirical analysis in this study adopts a data-driven approach to identifying
dynamical relationships among key variables relevant to multi-system sustainability
transitions. Rather than specifying a full structural model ex ante, we employ
sparse regression to infer a parsimonious set of governing relationships directly
from observed time series. This approach follows the logic of sparse identification
of nonlinear dynamical systems (SINDy), which has been proposed as a way to
discover interpretable dynamical equations from limited and noisy data
\citep{brunton2016discovering}. In the present study, this logic is implemented
through a sparse orthogonal regression formulation, in which the right-hand side
of the dynamical system is approximated using an orthogonal basis expansion.

The motivation for using an orthogonal basis is both practical and interpretive.
Sparse orthogonal regression has been shown to provide stable and compact
approximations of complex multivariate functions, with basis coefficients that
can be interpreted hierarchically and with reduced sensitivity to redundant or
highly correlated candidate terms \citep{roman2026utci}. At the same time, recent
work on dictionary-based dynamic-equation learning has emphasised that the
choice and conditioning of the candidate library can strongly affect sparse model
recovery, particularly when monomial libraries are ill-conditioned or when data
are noisy and limited \citep{feng2026illconditioning}. These considerations motivate
the use of an orthogonal representation in the present exploratory setting, while
also cautioning against over-interpreting individual coefficients as precise causal
estimates.

The starting point is the assumption that the temporal evolution of each variable
can be approximated by a low-dimensional dynamical system in which only a small
subset of candidate interactions is active. A library of potential explanatory
terms is constructed from the observed variables, including linear terms and
low-order nonlinear interactions, and represented through an orthogonal basis.
Sparse regression is then used to select the subset of terms that best explains
the observed rates of change, penalising model complexity to avoid overfitting.
This procedure yields a compact representation of the dominant dependencies
governing system evolution, while discarding statistically weak or redundant
relationships \citep{brunton2016discovering,feng2026illconditioning}.

The regression problem is linear in the unknown coefficients, but the resulting
dynamical model is not substantively linear: nonlinear dependencies are captured
through the basis functions and interaction terms included in the candidate
library. Moreover, all variables are normalised prior to estimation, ensuring that
coefficient magnitudes reflect relative importance rather than differences in
units or scale. The current implementation is application- and case-study-focused,
intended to demonstrate the usefulness of sparse orthogonal regression for
exploratory reconstruction of multi-system transition dynamics. The emphasis is
therefore on identifying structural relationships and sign-consistent influences,
not on precise numerical forecasting.

Conceptually, this approach aligns closely with the system dynamics tradition in
sustainability and transitions research, where feedback structures and causal
loops are central analytical objects \citep{papachristos2011system,holtz2015prospects}.
However, instead of specifying feedbacks based on expert judgement alone, the
present method infers candidate feedback structures empirically from data. The
resulting signed influence diagrams can thus be interpreted as empirically grounded
causal hypotheses, suitable for qualitative reasoning and scenario exploration
rather than deterministic prediction.

Within the broader sustainability transitions literature, this modelling strategy
addresses longstanding calls for approaches that can bridge empirical analysis
and dynamic systems thinking \cite{kohler2018modelling,kohler2019agenda,truffer2022perspective}.
Transitions are increasingly understood as the outcome of interacting social,
technological, economic, and environmental subsystems, characterised by feedbacks,
nonlinearities, and path dependence. Sparse data-driven dynamical models offer a
way to explore such interactions without imposing overly rigid theoretical
structures, while still retaining interpretability and analytical tractability.

Importantly, the method is used here as an exploratory and diagnostic tool rather
than a forecasting engine. As emphasised in recent debates on modelling and
governance, quantitative models should be understood as devices for structured
learning, sense-making, and deliberation, not as neutral or exhaustive
representations of reality \cite{stirling2023mind,schluter2024navigating,geels2022causality}.
The sparse regression framework adopted in this study is therefore intended to
support the identification of dominant dependencies and plausible feedback
structures that can inform subsequent qualitative interpretation, scenario
development, and participatory engagement, consistent with contemporary
multi-system transition research.

\begin{table}[t]
\centering
\footnotesize
\caption{Model variables, domains, units, and data sources}
\label{tab:variables}
\begin{tabular}{p{0.7cm} p{3.0cm} p{2.2cm} p{2.0cm} p{4.6cm}}
\hline
\textbf{ID} & \textbf{Variable} & \textbf{Domain} & \textbf{Unit} & \textbf{Data source} \\
\hline

V1 & Renewable energy share & Energy & \% &
Eurostat SHARES \cite{V1_Eurostat_REShare} \\

V2 & ETS emissions & Energy & MtCO\textsubscript{2}e &
EEA ETS database \cite{V2_EEA_ETS_Emissions} \\

V3 & Resource productivity & Economy & EUR/kg &
Eurostat env\_ac\_rp \cite{V3_Eurostat_ResourceProductivity} \\

V4 & Digitalisation & Technology & \% firms &
Eurostat ICT usage \cite{V4_Eurostat_Digitalisation} \\

V5 & Clean innovation & Technology & Number of patents &
OECD ENV-TECH patents \cite{V5_OECD_GreenPatents} \\

V6 & Policy stringency & Governance & Index &
OECD EPS index \cite{V6_OECD_PolicyStringency} \\

V7 & Transition finance & Economy & Index &
EEA green finance indicators \cite{V7_EEA_GreenBonds} \\

V8 & Social well-being & Social & Index &
Eurostat EU-SILC \cite{V8_Eurostat_Wellbeing} \\

V9 & Environmental stress & Environment & Index &
Eurostat WEI+ / climate \cite{V9_Eurostat_EnvironmentalStress} \\

V10 & System resilience & Governance & Index &
JRC resilience dashboards \cite{V10_JRC_Resilience} \\

\hline
\end{tabular}
\end{table}

\subsection*{Data and variables description}

To construct a prototype model that remains both tractable and meaningful, it is necessary to identify a coherent set of variables that represent the essential dimensions of socio-technical transition processes. The aim is to capture the key drivers, constraints, and interactions that shape medium-term system evolution, while avoiding unnecessary detail that would hinder calibration or reduce transparency. The selection process focuses on variables that are conceptually grounded in the transition literature, empirically measurable, and relevant for policy and strategic decision making. The resulting set of ten variables spans energy, industry, technology, finance, social conditions, governance, environmental stress, and system resilience. Together, these variables define the effective state space of the model, , see Table \ref{tab:variables}. They provide a structured representation of how different domains influence one another and how their combined trajectories shape transition pathways. Each variable is described in a separate subsection, including its conceptual motivation, potential data sources, and expected role within the system boundary adopted for this feasibility study. This structured approach ensures that the prototype model remains sufficiently comprehensive to reflect multi-dimensional transition dynamics, while still maintaining the simplicity required for medium-term calibration and interpretability.

\subsection{Renewable energy share}

The renewable energy share represents the proportion of total final energy consumption that originates from renewable energy sources, including solar, wind, hydro, geothermal, and sustainable biomass. Nuclear energy is not included in this indicator under current European statistical conventions. This variable captures a central structural dimension of energy system transformation and is widely used as a core indicator in sustainability transitions and decarbonisation research \citep{kohler2018modelling,kohler2019agenda}. Empirically, it is quantified using national energy balances, most commonly through the share of renewable energy in gross final energy consumption as defined under the EU Renewables Directives. Reliable data sources include Eurostat’s SHARES database and related energy statistics, which provide consistent annual data disaggregated by sector and energy carrier \citep{V1_Eurostat_REShare}.

From a modelling perspective, this variable plays a central role in shaping medium-term decarbonisation dynamics. Its evolution reflects both technological substitution processes and broader structural changes in energy demand, and it is influenced by policy instruments such as renewable support schemes, carbon pricing, grid integration measures, and fossil fuel phase-out regulations. In a system-level influence framework, the renewable energy share interacts with multiple domains: it contributes to reducing emissions-intensive lock-in, affects industrial energy use patterns, conditions transition finance requirements, and feeds back into social and political acceptance of climate policy. Given the path-dependent nature of energy system change, calibration relies on high-quality historical data rather than long-horizon optimisation-based projections. As a measurable, policy-relevant, and dynamically responsive indicator, renewable energy share constitutes a core state variable in the prototype model.

\subsection{Carbon-Intensive Capital Stock (Lock-In)}

Carbon-intensive capital stock represents the installed base of high-emission physical assets that structure the inertia of the socio-technical system. This includes fossil-based power plants, energy-intensive industrial installations, internal combustion vehicle fleets, fossil fuel infrastructure, building stock with poor energy performance, and other long-lived assets whose operational characteristics embed emissions-intensive pathways. In sustainability transitions research, such installed asset bases are widely recognised as a key mechanism of path dependence and lock-in, shaping the speed and feasibility of structural change \citep{kohler2019agenda,truffer2022perspective}.

Empirically, direct reconstruction of the full capital stock is difficult at aggregate scales and over limited observation windows. In this feasibility study, carbon-intensive capital stock is therefore proxied by verified emissions from large stationary installations covered by the EU Emissions Trading System (ETS), focusing on power and industry. This provides a consistent, policy-relevant indicator of the extent and utilisation of emissions-intensive infrastructure, and offers a transparent empirical anchor for the model’s ``carbon lock-in'' dimension \citep{V2_EEA_ETS_Emissions}. While emissions are not a one-to-one measure of installed capacity, they reflect the operational imprint of the incumbent asset base and are suitable for capturing medium-term inertia in a compact dynamical prototype.

From a socio-technical modelling perspective, emissions-intensive capital stock is a central source of path dependence. High levels of embedded fossil infrastructure constrain the pace of transition pathways, raise adjustment challenges, and delay the diffusion of low-carbon alternatives. Lock-in also interacts with governance and political-economy dynamics, since incumbency and sunk investments condition which interventions are regarded as feasible or legitimate \citep{geels2022causality,turnheim2015evaluating}.

Calibration in this study focuses on reproducing observed ETS emissions dynamics rather than reconstructing asset inventories. Medium-term evolution reflects combined effects of regulatory tightening, fuel switching, technological substitution, and changes in utilisation rates. As a slow-moving but highly consequential system variable, carbon-intensive capital stock anchors the medium-term dynamic constraints in the prototype model and provides an empirically grounded representation of carbon lock-in over the analysis window.

\subsection{Industrial Energy \& Material Intensity}

Industrial energy and material intensity captures the amount of primary materials required to generate economic output in industrial systems. It is widely used in sustainability transitions research because it reflects both technological efficiency improvements and deeper structural shifts in production, consumption, and value creation \citep{kohler2019agenda,truffer2022perspective}. In this study, the concept is operationalised using resource productivity, defined as gross domestic product per unit of domestic material consumption, which provides an aggregate indicator of material efficiency at the economy–industry interface.

Empirically, resource productivity integrates information on material extraction, trade, and use relative to economic output, and is commonly applied in European sustainability assessments as a proxy for material and industrial intensity. The indicator is constructed from economy-wide material flow accounts and national accounts data and is available as a consistent long-term time series for the European Union \citep{V3_Eurostat_ResourceProductivity}. While it does not resolve individual industrial subsectors, it captures system-level efficiency trends associated with structural change, technological upgrading, and shifts in production patterns.

Within the modelling framework, this variable provides a quantitative link between economic activity and environmental pressure. Rising resource productivity signals decoupling tendencies, including process efficiency gains, circular economy practices, and changes in industrial composition. Conversely, stagnating or declining productivity indicates persistent material intensity and continued pressure on natural resources. Rather than projecting detailed sectoral pathways, the prototype model uses the indicator to represent aggregate efficiency dynamics over the period of analysis. As a slowly evolving but structurally important variable, industrial energy and material intensity plays a critical role in shaping the medium-term evolution of provisioning systems within the model.

\subsection{Digitalisation Intensity of the Economy}

Digitalisation intensity reflects the degree to which economic activities and industrial processes rely on digital technologies and data-driven applications. It captures the diffusion of information and communication technologies (ICT) that support coordination, automation, and market integration across firms and sectors. In this study, digitalisation intensity is proxied by the share of enterprises engaging in e-commerce activities, which provides a consistent and widely used indicator of digital adoption in the productive economy.

Empirically, the indicator is drawn from Eurostat statistics on ICT usage in enterprises, which track the proportion of firms with e-commerce sales across the European Union \citep{V4_Eurostat_Digitalisation}. While this measure does not capture the full breadth of advanced digital technologies, it reflects the penetration of digital business models and online transaction capabilities, which are closely associated with broader digital transformation processes. As such, it offers a parsimonious and empirically robust proxy for digitalisation dynamics over the available time horizon.

From a modelling standpoint, digitalisation intensity acts primarily as an enabling variable in socio-technical transitions. Higher levels of digital adoption can support reductions in energy and material intensity through improved coordination, logistics optimisation, and process efficiency, while also facilitating innovation diffusion and market access for emerging technologies. The empirical time series allows the identification of steady diffusion dynamics and periods of acceleration or saturation in digital adoption across the economy. In the model, this variable is used to capture the extent to which digital practices are embedded in economic activity, without resolving individual technologies. Digitalisation intensity therefore enters as a structural condition that shapes coordination capacity, innovation spillovers, and system responsiveness over medium-term transition horizons.

\subsection{Clean Innovation \& Diffusion Rate}

Clean innovation captures the development of low-carbon, resource-efficient, and sustainability-oriented technologies that enable structural change across energy, industrial, and provisioning systems. In this study, innovation is operationalised through counts of environment-related patents, which provide a widely used proxy for the direction and intensity of technological effort devoted to environmental objectives. Patent-based indicators are commonly employed in transitions research to trace innovation dynamics, knowledge accumulation, and shifts in technological search processes \citep{plank2021doing,knobloch2016behavioural}.

Empirically, the variable is constructed using OECD ENV-TECH patent data, which classify patent families related to environmental and climate-relevant technologies based on internationally accepted criteria \citep{V5_OECD_GreenPatents}. Patent counts do not directly measure deployment or diffusion, but they reflect upstream innovation activity and expectations about future technological relevance. As such, they are well suited for capturing medium-term innovation dynamics in a compact modelling framework, particularly when consistent deployment data are unavailable across sectors and time.

Within socio-technical transition modelling, clean innovation functions as a key enabling process that shapes the feasibility and direction of transformation pathways. Sustained innovation activity supports cost reductions, performance improvements, and the emergence of viable alternatives to incumbent technologies, thereby reducing dependence on emissions-intensive capital and expanding the space of politically and economically feasible policy options \citep{holtz2015prospects,geels2022causality}. 

The historical patent time series allows identification of phases of acceleration, stagnation, and recovery in innovation activity, reflecting broader economic cycles and policy signals. In the modelling, we envision clean innovation as a dynamic state variable that modulates the system’s adaptive capacity over the medium term. Rather than predicting specific technologies, it represents the structural role of innovation in enabling or constraining transition pathways across interconnected systems.

\subsection{Policy Stringency \& Credibility}

Policy stringency and credibility captures the overall strength, consistency, and perceived reliability of public policies aimed at steering socio-technical change. The variable reflects multiple dimensions of governance, including the ambition of environmental and climate-related targets, the scope and design of regulatory instruments, enforcement practices, and the temporal stability of policy commitments. Empirically, policy stringency can be proxied using composite indicators such as the OECD Environmental Policy Stringency (EPS) Index, complemented where relevant by information on carbon pricing instruments, regulatory standards, and sector-specific policy measures \citep{V6_OECD_PolicyStringency,galeotti2020measuring}. While credibility is more difficult to quantify directly, it can be approximated through the persistence of policy frameworks over time, consistency between stated targets and observed implementation, and institutional continuity reflected in historical policy trajectories.

In the modelling framework, policy stringency and credibility is treated as a governance state variable that conditions incentives, expectations, and strategic behaviour across interconnected systems. Rather than pre-specifying its precise effects, the model allows policy stringency to interact with other variables—such as innovation activity, capital turnover, industrial intensity, and financial conditions—through empirically calibrated feedbacks. In this sense, policy stringency may shape the pace and direction of system change by influencing investment signals, technology adoption, and the retirement of carbon-intensive assets, while the magnitude and relevance of these interactions are inferred from the data rather than imposed a priori. At the same time, policy evolution itself may respond to pressures arising from social outcomes, environmental stress, and structural economic constraints, reflecting the endogenous nature of governance in transition processes.

Forward-looking assumptions may be informed by existing EU-level policy frameworks and long-term strategies, but are used primarily to explore qualitative system responses rather than to encode deterministic pathways. As a high-leverage but institutionally embedded variable, policy stringency and credibility provides an essential governance channel through which medium-term transition dynamics can be examined within the prototype model.

\subsection{Transition Finance Conditions}

Transition finance conditions capture the availability, cost, and allocation of financial resources relevant for the reorientation of investment toward low-carbon and resource-efficient activities. The variable reflects structural features of the financial system, including the scale of sustainable finance flows, relative financing conditions for low- versus high-carbon activities, and the extent to which capital markets and public financial instruments support transition-oriented investments. Empirically, transition finance can be proxied using indicators such as green bond issuance volumes, sustainable finance indices, and aggregate measures of low-carbon investment mobilisation at the European level \citep{V7_EEA_GreenBonds,smolenska2025european}. These indicators provide an observable signal of how financial systems engage with transition objectives, without presupposing specific allocation mechanisms at the project level.

Within the modelling framework, transition finance conditions act as a mediating variable that conditions the pace and direction of capital reallocation across interconnected systems. Rather than assuming that finance directly drives specific outcomes, the model allows financial conditions to interact with other system variables—such as innovation activity, capital stock dynamics, policy signals, and stress factors—through empirically inferred feedbacks. In this sense, favourable transition finance conditions may coincide with faster deployment of low-carbon technologies, greater scope for retrofitting or asset turnover, and enhanced system stability, while constrained or misaligned financial conditions may be associated with persistence of incumbent infrastructures and delayed adjustment. The strength and sign of these relationships are not imposed ex ante, but emerge from the regression-based identification of the system’s dynamical structure.

The historical time series captures the expansion of sustainable finance instruments and their temporal co-movement with other transition-relevant variables. This enables the model to represent periods of rapid financial mobilisation as well as phases of stagnation or volatility. The transition finance conditions is cross-cutting economic variable and provides an essential link between governance signals, technological change, and the material evolution of provisioning systems in the prototype model.

\subsection{Social Well-Being \& Just Transition Pressure}

Social well-being captures subjective assessments of quality of life that are relevant for the social legitimacy and political sustainability of socio-technical transitions. In this study, the variable is operationalised using overall life satisfaction, as measured through EU-SILC survey data \citep{V8_Eurostat_Wellbeing}. Life satisfaction provides a concise summary indicator that reflects how individuals perceive their material conditions, economic security, social environment, and future prospects, without disaggregating these dimensions into separate components.

Within the modelling framework, social well-being is treated as an aggregate social-state variable that conditions the feasibility of transition pathways rather than as a direct policy outcome. Changes in life satisfaction may coincide with shifts in economic conditions, energy affordability, employment prospects, or exposure to environmental stress, all of which are salient during periods of structural transformation. The model does not impose a specific mechanism linking well-being to governance or economic variables; instead, it allows the direction and strength of interactions between social well-being, policy stringency, transition finance, and environmental stress to emerge from the empirical fit.

From a modelling perspective, life satisfaction may respond to disruptions associated with industrial restructuring, price volatility, or environmental pressures, while also interacting with governance capacity and institutional stability. This positioning is consistent with sustainability transitions literature that emphasises the role of societal legitimacy and perceived quality of life as mediating factors in the durability of transition strategies \citep{scoones2020transformations,giang2024equity}. Thus, the variable enables the model to capture periods of social strain as well as relative stability. As a parsimonious yet informative indicator, overall life satisfaction provides a tractable representation of social conditions relevant to medium-term transition dynamics.

\subsection{Resource \& Environmental Stress}

Resource and environmental stress captures biophysical pressures that constrain socio-technical systems and shape the conditions under which transitions unfold. In this study, the variable is operationalised using a composite representation based primarily on the Water Exploitation Index (WEI+) and related climate- and environment-related stress indicators at the EU level \citep{V9_Eurostat_EnvironmentalStress}. These indicators reflect pressures associated with water scarcity and variability, climatic extremes, and broader environmental constraints that affect economic activity and provisioning systems. As such, the variable represents an aggregate measure of environmental stress rather than a comprehensive accounting of all ecological dimensions.

Within the modelling framework, environmental stress is treated as a system-level condition that interacts with economic, social, and technological dynamics over the medium term. Elevated stress levels may coincide with reduced industrial productivity, disruptions to supply chains, increased exposure to climate-related hazards, or heightened pressure on critical infrastructures, while periods of lower stress may ease operational constraints. The model does not impose a predefined causal pathway; instead, it allows the direction and strength of interactions between environmental stress and other variables—such as emissions, resource productivity, social well-being, and transition finance—to emerge from the empirical fit.

Environmental stress can respond to changes in emissions-intensive activity and material use, while also feeding back into socio-economic conditions through impacts on production, prices, and exposure to environmental risks. This positioning is consistent with nexus-oriented and socio-ecological systems perspectives that emphasise environmental stress as a cross-cutting constraint linking ecological conditions to economic and social outcomes \citep{allouche2024nexus,schluter2024navigating}. The index allows to capture both gradual pressures and episodic variability over the observation window. It is an aggregated environmental constraint, but the resource and environmental stress variables provides an empirically grounded representation of biophysical limits relevant for analysing medium-term transition dynamics.

\subsection{Resilience of Provisioning Systems}

In this study, the resilience of provisioning systems is conceptualised and operationalised
through the lens of \emph{financial integration}. Rather than capturing physical robustness
or infrastructural redundancy directly, this variable reflects the capacity of financial
systems to sustain, stabilise, and reallocate investment flows that underpin essential
provisioning systems such as energy, industry, mobility, and related infrastructures.
Financial integration is therefore treated as a conditioning factor that shapes the ability
of socio-technical systems to absorb disturbances and maintain functionality under
conditions of stress.

Financially integrated systems are characterised by diversified funding sources, access to
long-term capital, and risk-sharing mechanisms across sectors and jurisdictions. These
features can enhance adaptive capacity by reducing exposure to abrupt funding constraints,
pro-cyclical investment dynamics, and fragmented responses to external shocks. Conversely,
low levels of financial integration may amplify vulnerability by constraining adaptive
investment, increasing volatility, and propagating disturbances across interconnected
systems.

Empirically, resilience is proxied using composite indicators that capture the degree of
financial integration supporting key provisioning systems, drawing on the Joint Research
Centre’s resilience dashboards and related EU-level indicators \citep{V10_JRC_Resilience}.
These indicators reflect structural features of financial systems that influence continuity
and adjustment in the face of environmental stress, economic disruption, and policy-driven
transition pressures. While such proxies do not measure physical resilience directly, they
provide an empirically grounded approximation of the financial conditions that enable or
constrain system-wide adaptability.

Within the modelling framework, resilience functions primarily as a stabilising modifier
rather than as a direct driver of transformation. Higher financial integration is associated
with a dampening of shock transmission by sustaining investment flows and smoothing
adjustment pathways, thereby supporting service continuity and social stability. Lower
integration, by contrast, may coincide with heightened stress and reduced coordination
capacity across systems.

Understood in this way, resilience, interpreted as financial integration, shapes the
robustness and coherence of medium-term sustainability transition pathways without
prescribing their direction or outcomes.

\section{Results and discussion}

This section compares empirical behaviour with two projection approaches: the 
learned dynamical system (DS) model and a simple univariate linear 
extrapolation. Each figure reports observed data, the DS reconstruction over the 
calibration period, and forward projections over a 20 year horizon. The linear 
projection represents trend continuation without acknowledging interactions among 
variables. The DS model, in contrast, incorporates inferred dependencies among 
variables. This allows us to see where projections remain close to linear 
continuation and where cross-system conditions become relevant for future 
pathways.

\begin{figure}[t]
  \centering
  \includegraphics[width=\textwidth]{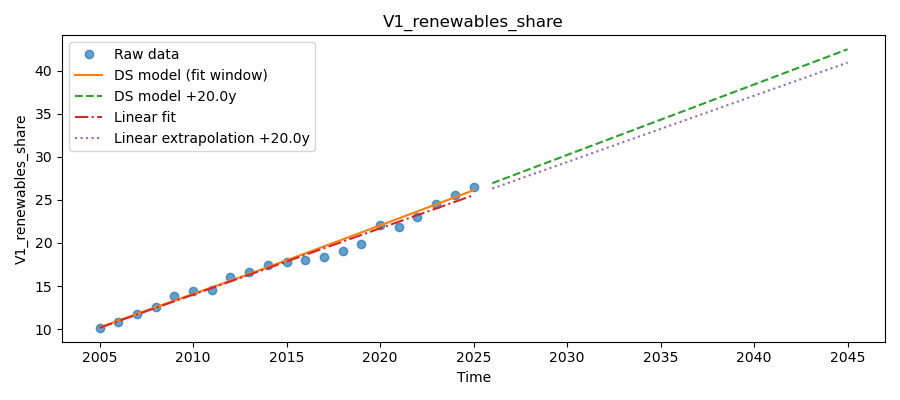}
  \caption{Empirical time series and dynamical system (DS) model fit for the share of renewables (V1\_renewables\_share). The indicator is the percentage of renewable energy in total energy use (EU-level proxy), shown together with the DS reconstruction and +20-year projection, and a simple linear fit with extrapolation.}
  \label{fig:v1_renewables}
\end{figure}

\subsection{V1 Renewables Share}

Figure~\ref{fig:v1_renewables} reports the share of renewable energy in total final energy use (Eurostat renewable share). The DS reconstruction closely reproduces the observed historical growth pattern, matching both the level and slope of the empirical series. Forward projections remain close to the linear extrapolation, indicating that recent expansion dynamics are likely to persist over the medium term. Importantly, the DS formulation reaches this outcome while explicitly embedding cross-system interactions and nonlinear structure, showing that continued growth is robust to the inclusion of feedbacks from policy, innovation, finance, and environmental conditions. The slightly higher DS trajectory reflects weak but persistent reinforcement rather than a qualitative regime shift. The close alignment between the DS and linear projections therefore acts as a consistency check: renewable deployment over this horizon appears structurally stable, not an artefact of simplistic trend extrapolation.

\begin{figure}[t]
  \centering
  \includegraphics[width=\textwidth]{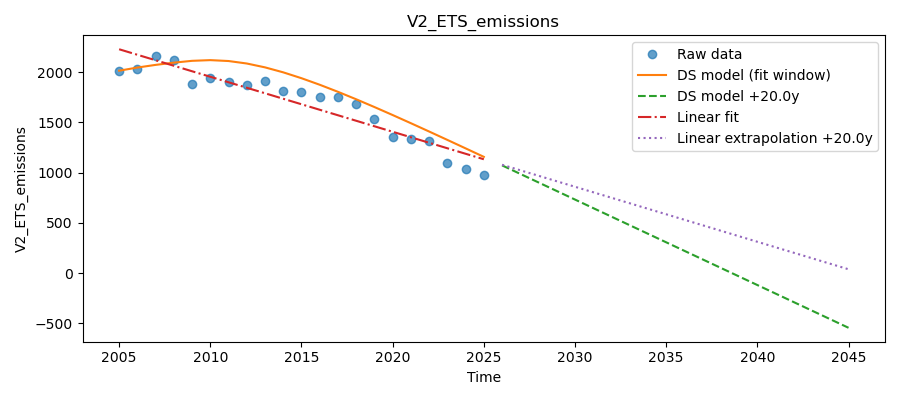}
  \caption{ETS-regulated emissions (V2\_ETS\_emissions): verified greenhouse gas emissions from sectors covered by the EU Emissions Trading System (EU ETS), used as a proxy for industrial climate pressure. The figure compares the raw series with the DS model fit and +20-year projection, and with the linear regression fit and its extrapolation.}
  \label{fig:v2_ets}
\end{figure}

\subsection{V2 ETS-Regulated Emissions}

Figure~\ref{fig:v2_ets} shows verified greenhouse gas emissions from sectors covered by the EU Emissions Trading System (ETS). Both the linear and the dynamical system (DS) models reproduce the historical downward trend over the estimation window. However, their projections diverge markedly beyond the observed period.

The linear extrapolation assumes that the past rate of decline continues smoothly into the future, resulting in a gradual and mechanically persistent reduction in emissions. By contrast, the DS projection exhibits a more pronounced decline, reflecting the interaction of emissions with other system variables rather than a simple continuation of historical momentum. In the inferred model structure, emissions are endogenously linked to environmental stress, financial dynamics, and productivity-related feedbacks, which together generate a non-linear acceleration of reductions under favourable structural conditions.

This behaviour indicates that deeper emissions cuts are not driven by time alone, but by the alignment of broader system dynamics. When reinforcing conditions are present, emissions reductions can proceed faster than suggested by linear trend extrapolation. Conversely, the model structure also implies that such outcomes are conditional rather than guaranteed, as changes in stress, finance, or productivity could weaken or reverse the trajectory.

The DS projection should therefore not be interpreted as an optimistic forecast, but as a structurally contingent pathway that highlights the potential for accelerated decarbonisation when system-level interactions are aligned. The linear extrapolation, while smooth and intuitive, obscures these conditional dynamics by assuming a fixed slope that is insensitive to changes in the wider socio-economic and institutional context.

\newpage

\begin{figure}[t]
  \centering
  \includegraphics[width=\textwidth]{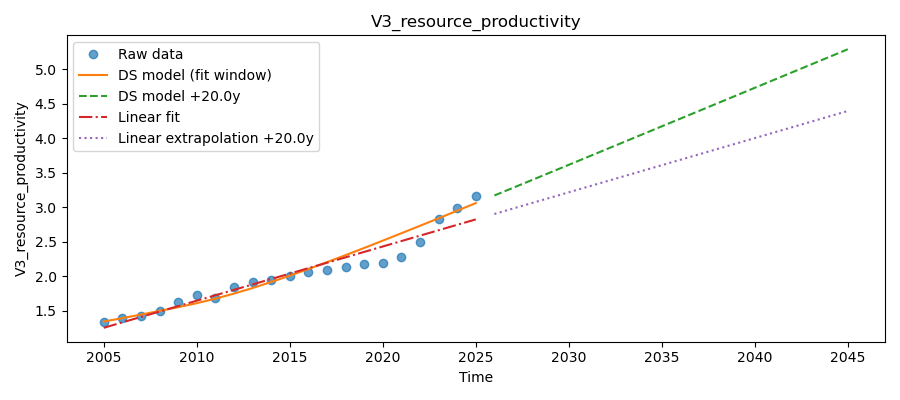}
  \caption{Resource productivity (V3\_resource\_productivity), proxied by GDP per unit of domestic material consumption. The plot shows observed data, DS model reconstruction and forward simulation, and the linear trend with extrapolation.}
  \label{fig:v3_resource_productivity}
\end{figure}

\begin{figure}[t]
  \centering
  \includegraphics[width=\textwidth]{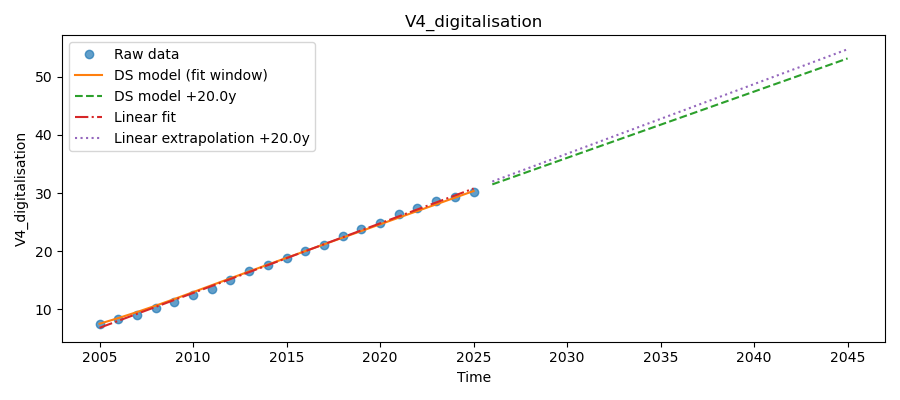}
  \caption{Digitalisation (V4\_digitalisation), measured as the share of enterprises with e-sales in total enterprises (Eurostat e-commerce indicator). Raw data are contrasted with the DS model fit and +20-year projection, as well as the linear fit and extrapolated trend.}
  \label{fig:v4_digitalisation}
\end{figure}

\begin{figure}[t]
  \centering
  \includegraphics[width=\textwidth]{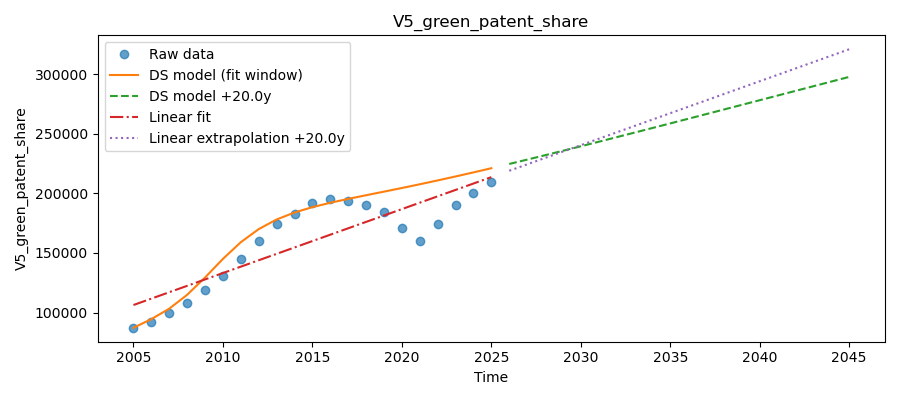}
  \caption{Green patent share (V5\_green\_patent\_share), representing the proportion of environment- or climate-related patents in total patenting activity (EU proxy for low-carbon innovation). The figure shows the empirical series together with the DS model reconstruction and projection and the linear regression baseline.}
  \label{fig:v5_green_patents}
\end{figure}

\begin{figure}[t]
  \centering
  \includegraphics[width=\textwidth]{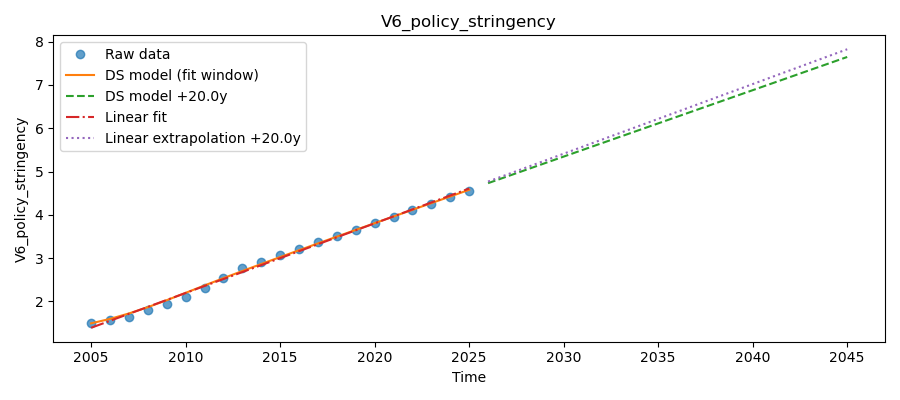}
  \caption{Environmental policy stringency (V6\_policy\_stringency), based on the OECD Environmental Policy Stringency (EPS) index averaged for European countries (0--6 scale). The plot displays the observed evolution, the DS model fit and +20-year projection, and the corresponding linear fit and extrapolation.}
  \label{fig:v6_policy_stringency}
\end{figure}

\begin{figure}[t]
  \centering
  \includegraphics[width=\textwidth]{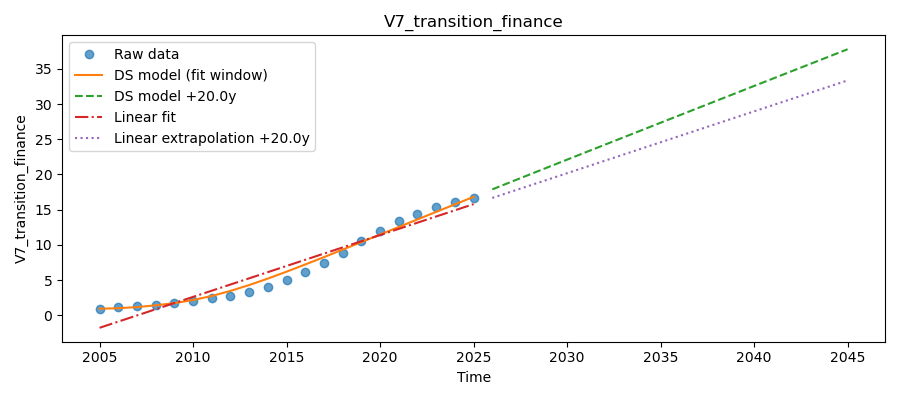}
  \caption{Transition finance (V7\_transition\_finance), proxied by green bond issuance as a percentage of total bond issuance (all issuers, EU--27). The empirical series is compared with the DS model dynamics and the linear trend, including a +20-year extrapolation.}
  \label{fig:v7_transition_finance}
\end{figure}

\begin{figure}[t]
  \centering
  \includegraphics[width=\textwidth]{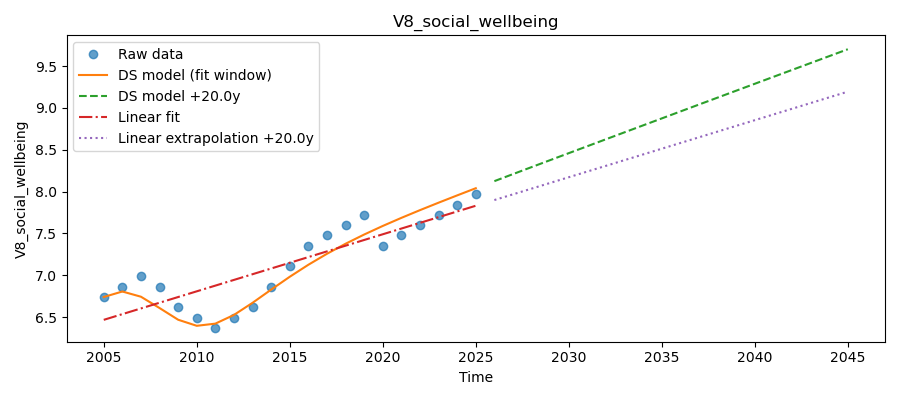}
  \caption{Social wellbeing (V8\_social\_wellbeing), using average overall life satisfaction scores (0--10 scale) from European survey data as a quality-of-life proxy. The figure compares DS model fit and projection with raw observations and the linear regression baseline.}
  \label{fig:v8_social_wellbeing}
\end{figure}

\begin{figure}[t]
  \centering
  \includegraphics[width=\textwidth]{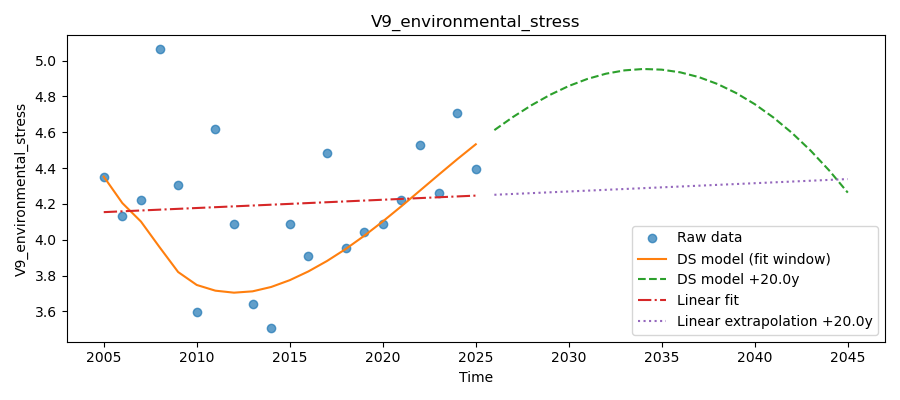}
  \caption{Environmental stress (V9_environmental_stress), measured using the Water Exploitation Index (WEI+), which captures pressure on renewable freshwater resources from abstraction relative to availability (see Methods). The figure compares the observed WEI+ time series with the dynamical system (DS) reconstruction and forward simulation, alongside a linear fit and its extrapolation.}
  \label{fig:v9_environmental_stress}
\end{figure}

\begin{figure}[t]
  \centering
  \includegraphics[width=\textwidth]{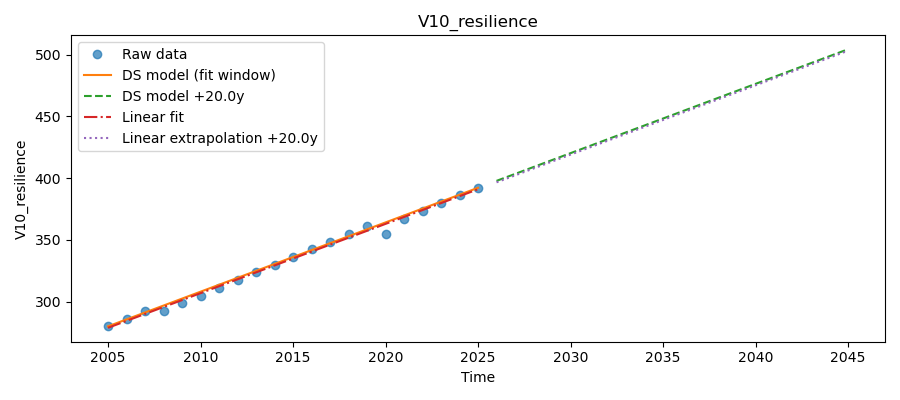}
  \caption{System resilience (V10\_resilience), proxied by EU financial integration. The indicator measures the sum of external financial assets and external financial liabilities (both intra--EU and extra--EU), expressed as a percentage of GDP. Higher levels indicate greater capacity for risk-sharing, cross-border capital absorption, and shock buffering within the financial system. The figure shows raw historical data compared to the DS model dynamics (fit window and +20-year projection) and to the linear regression fit and extrapolation.}
  \label{fig:v10_resilience}
\end{figure}

\subsection{V3 Resource Productivity}

Figure~\ref{fig:v3_resource_productivity} reports resource productivity, measured as GDP per unit of domestic material consumption (Eurostat). Both modelling approaches reproduce the gradual historical increase observed over the estimation period, indicating a persistent improvement in material efficiency across the European economy.

Beyond the fit window, the projections diverge. The linear extrapolation assumes that historical gains continue at a constant rate, effectively treating improvements in resource productivity as a steady, time-driven process. By contrast, the dynamical system (DS) projection exhibits a slightly faster increase, indicating an acceleration of productivity gains relative to the linear continuation.

This divergence reflects the DS model’s capacity to internalise reinforcing structural mechanisms rather than extrapolating past trends mechanically. Improvements in resource productivity can compound through coordinated technological upgrading, process optimisation, and organisational learning, especially when supported by innovation dynamics and investment patterns. The DS structure allows such reinforcing effects to emerge endogenously, producing a convex trajectory without imposing exogenous growth assumptions.

\subsection{V4 Digitalisation}

Figure~\ref{fig:v4_digitalisation} shows the share of enterprises conducting e-sales, used as an indicator of business digitalisation (Eurostat). Both modelling approaches reproduce the historical trajectory closely, reflecting the steady and near-linear diffusion of digital practices across firms over the observation period.

Beyond the estimation window, the linear extrapolation assumes a continued constant-rate expansion. The dynamical system (DS) projection follows a very similar trajectory but increases marginally more slowly. This difference is modest and does not indicate saturation or reversal; rather, it reflects the DS model’s sensitivity to structural conditions that can modulate the pace of diffusion without interrupting its overall upward trend.

Importantly, the DS representation does not suppress ongoing adoption dynamics. Instead, it allows digitalisation growth to remain contingent on complementary factors such as organisational capacity, institutional stability, and innovation conditions, without imposing sharp nonlinear transitions or diffusion ceilings. The close alignment between the historical fit and the projected trajectory suggests that endogenous feedbacks are present but not exaggerated. Overall, the projection indicates that digitalisation continues to expand in a stable and sustained manner.

\subsection{V5 Green Patent Share (Innovation)}

Figure~\ref{fig:v5_green_patents} shows the evolution of green patenting activity, measured using environment-related patent classifications. The DS reconstruction closely follows the underlying empirical trajectory while smoothing short-term fluctuations in the data, including the temporary decline observed around 2020.

In projection, the DS model continues the upward trend but at a slightly lower rate than the linear extrapolation. This reflects that the inferred dynamics capture medium-term structural tendencies rather than extrapolating short-lived shocks or cyclical disturbances. By contrast, the linear fit directly propagates the influence of recent observations without distinguishing between transient disruptions and persistent drivers.

The absence of a pronounced pandemic-related downturn in the DS projection highlights its role as a structural representation of innovation dynamics rather than a short-term forecasting tool. In this sense, the DS model extracts the dominant systemic trend in clean innovation, providing a more stable basis for analysing longer-run transition dynamics than approaches that mechanically extend recent fluctuations.

\subsection{V6 Policy Stringency}

Figure~\ref{fig:v6_policy_stringency} reports the OECD Environmental Policy Stringency (EPS) index averaged across European countries. The historical trajectory is close to linear, reflecting the cumulative and incremental nature of regulatory change. Both the DS reconstruction and the linear fit reproduce this pattern almost exactly, and their forward projections remain closely aligned.

This convergence indicates that, over the observed period, policy stringency behaves as a slow-moving structural variable rather than a dynamically responsive one. Changes in regulatory ambition appear to be driven primarily by institutional processes, negotiated targets, and administrative sequencing, rather than by fast feedbacks from economic or environmental conditions. The DS model therefore does not introduce additional curvature or endogenous acceleration, but instead preserves the empirically observed governance rhythm.

The result is informative rather than trivial. It suggests that within this dataset, policy stringency provides a stable background condition that shapes other system dynamics without itself responding strongly on medium-term timescales. In this sense, policy operates as a conditioning framework rather than a leading dynamical driver. The close agreement between the two projections reinforces confidence that the DS formulation respects this structural property instead of imposing artificial feedback behaviour.

\subsection{V7 Transition Finance}

Figure~\ref{fig:v7_transition_finance} shows the share of green bond issuance as a fraction of total bond issuance, drawing on Climate Bonds Initiative and ECB data. Both models reproduce the limited growth observed prior to around 2016, followed by a phase of rapid expansion. The DS projection rises more strongly than the linear extrapolation, reflecting endogenous reinforcement once financial infrastructure, policy credibility, and market depth become established.

This acceleration is not imposed ex ante but emerges from the fitted system dynamics, consistent with empirical evidence that transition finance scales nonlinearly as coordination and signalling mechanisms mature. By contrast, the linear extrapolation assumes constant marginal growth and therefore understates the potential for cumulative financial mobilisation.

Among the variables considered, transition finance is one where cross-system interdependencies are particularly salient: capital allocation responds to institutional alignment and confidence effects rather than to time-driven continuation alone. The DS projection thus highlights how financial conditions can shift regime once enabling structures are in place.

\subsection{V8 Social Wellbeing}

Figure~\ref{fig:v8_social_wellbeing} uses average overall life satisfaction from European survey data (Eurofound / ESS indicators). Both models reproduce the gradual upward trend observed over the historical period, including the temporary decline around the late 2000s and early 2010s. The DS reconstruction captures this dip more explicitly than the linear fit, reflecting sensitivity to medium-term systemic disruptions rather than smoothing them away.

In the forward projection, the DS trajectory rises more strongly than the linear continuation, indicating a sustained recovery and consolidation of well-being gains once stabilising conditions are re-established. This behaviour suggests that improvements in social well-being are not purely time-driven but are reinforced when broader socio-economic and institutional conditions align.

The linear extrapolation, by contrast, assumes a constant marginal increase and does not reflect the possibility of post-shock rebound dynamics. The DS result does not imply guaranteed improvement, but highlights that well-being trajectories are shaped by systemic context rather than momentum alone. The close historical fit supports interpreting the projection as a conditional continuation of observed recovery patterns rather than an artefact of model structure.

\subsection{V9 Environmental Stress}

Figure~\ref{fig:v9_environmental_stress} reports environmental stress measured using the Water Exploitation Index (WEI+), which captures pressure on renewable freshwater resources arising from abstraction relative to availability. The historical series exhibits substantial variability rather than a monotonic trend, reflecting the sensitivity of water stress to climatic conditions, sectoral demand, and episodic shocks such as drought years.

The linear fit and extrapolation impose a smooth and gradually increasing trajectory, implicitly assuming that water stress accumulates steadily over time. By contrast, the DS reconstruction captures a non-linear profile within the historical window, including periods of declining stress followed by renewed increases. This behaviour is consistent with the hydrological nature of the indicator, where stress responds to interacting drivers rather than simple time dependence.

In the forward projection, the DS trajectory rises into the medium term before levelling off and partially declining toward the end of the horizon. This pattern does not imply resolution of water stress, but indicates that pressure on water resources is conditionally mediated by system-wide adjustments, including changes in industrial intensity, investment patterns, and adaptive capacity. The projected stress remains elevated throughout, but does not follow an unchecked upward path.

This contrast highlights the added value of the dynamical system formulation. Water stress is not well represented as a linear extrapolation of past values, as it reflects feedbacks between demand, infrastructure, economic activity, and adaptation rather than cumulative accumulation. The DS projection therefore provides a structurally coherent representation of medium-term water stress dynamics, capturing conditional stabilisation effects that are obscured by trend-based approaches.

\subsection{V10 Resilience (Financial Integration)}

Figure~\ref{fig:v10_resilience} reports system resilience proxied by financial integration, measured as the sum of external financial assets and liabilities relative to GDP (ECB Financial Integration indicators). The historical series exhibits a steady upward trajectory, reflecting the gradual deepening of cross-border financial linkages rather than episodic shocks or regime shifts. Both the dynamical system (DS) and linear models reproduce this behaviour closely over the calibration window.

The forward projections remain similar across modelling approaches, indicating that financial integration evolves as a slow-moving structural condition rather than as a source of strong endogenous feedbacks within the analysed period. This is consistent with empirical interpretations of financial integration as enabling risk sharing, liquidity provision, and investment continuity, rather than generating short-term volatility.

The close correspondence between the DS reconstruction and observed data supports the interpretation of resilience as a stabilising background capacity in the system. Its role in the model is therefore not to drive transformation directly, but to condition the system’s ability to absorb stress and sustain coordinated adjustment. The similarity between DS and linear projections is itself informative, suggesting that resilience contributes to robustness and continuity rather than to nonlinear amplification of transition dynamics.

\section{Discussion}

This section interprets the sparse dynamical system identified in the preceding
analysis, with the aim of situating the inferred causal structure within the
broader literature on multi-system transitions (MST). The results are not
presented as predictive or prescriptive; rather, they are used to elucidate
persistent dependencies and feedbacks that shape the joint evolution of social,
technological, economic, and environmental subsystems. In this sense, the model
serves as a diagnostic device that reveals structurally relevant interactions,
while remaining agnostic about optimal pathways or policy targets.

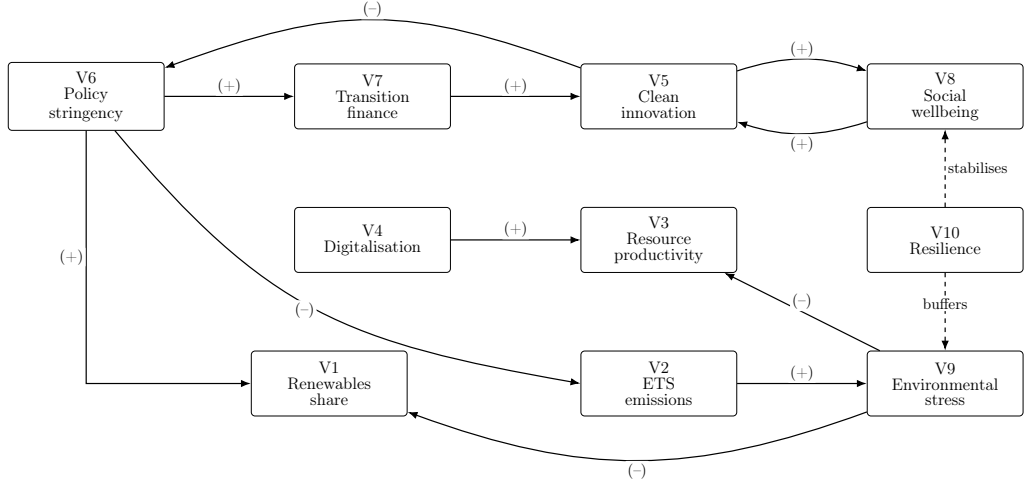
\begin{figure}[t]
\centering
\resizebox{0.98\linewidth}{!}{%
\begin{tikzpicture}[
  font=\small,
  >={Latex[length=3mm]},
  node distance=18mm and 30mm,
  box/.style={
    rounded corners=2.5pt, draw, align=center,
    inner sep=6pt, minimum width=36mm, minimum height=15mm
  },
  arr/.style={-Latex, thick},
  darr/.style={-Latex, thick, dashed},
  lab/.style={font=\footnotesize, fill=white, inner sep=1.5pt, rounded corners=1pt}
]

\node[box] (V6) {\shortstack{V6\\Policy\\stringency}};
\node[box, right=of V6] (V7) {\shortstack{V7\\Transition\\finance}};
\node[box, right=of V7] (V5) {\shortstack{V5\\Clean\\innovation}};
\node[box, right=of V5] (V8) {\shortstack{V8\\Social\\wellbeing}};

\node[box, below=of V7] (V4) {\shortstack{V4\\Digitalisation}};
\node[box, right=of V4] (V3) {\shortstack{V3\\Resource\\productivity}};
\node[box, right=of V3] (V10) {\shortstack{V10\\Resilience}};

\node[box, below=of V3] (V2) {\shortstack{V2\\ETS\\emissions}};
\node[box, right=of V2] (V9) {\shortstack{V9\\Environmental\\stress}};

\node[box, left=of V2, xshift=-10mm] (V1) {\shortstack{V1\\Renewables\\share}};


\draw[arr] (V6) -- node[lab, above] {(+)} (V7);

\draw[arr] (V6.south) |- node[lab, pos=0.25, left] {(+)} (V1.west);

\draw[arr, out=-50, in=170, looseness=1.15]
  (V6) to node[lab, left] {(--)} (V2.west);

\draw[arr] (V7) -- node[lab, above] {(+)} (V5);

\draw[arr, bend left=18] (V5) to node[lab, above] {(+)} (V8);
\draw[arr, bend left=18] (V8) to node[lab, below] {(+)} (V5);

\draw[arr, out=160, in=20, looseness=1.15]
  (V5) to node[lab, above] {(--)} (V6);

\draw[arr] (V4) -- node[lab, above] {(+)} (V3);

\draw[arr] (V2) -- node[lab, above] {(+)} (V9);

\draw[arr, out=200, in=340, looseness=1.05]
  (V9) to node[lab, below] {(--)} (V1);

\draw[arr] (V9) -- node[lab, above] {(--)} (V3);

\draw[darr] (V10) -- node[lab, above] {buffers} (V9);
\draw[darr, out=90, in=270, looseness=1.15]
  (V10) to node[lab, right] {stabilises} (V8);

\end{tikzpicture}%
}
\caption{Causal loop diagram (CLD) representing the structural dependencies encoded in the prototype dynamical system. Solid arrows denote signed influences between state variables, while dashed arrows indicate buffering or stabilising capacities. The CLD highlights cross-system feedbacks shaping medium-term transition dynamics across energy, innovation, finance, governance, social well-being, and environmental stress, and serves as an interpretive complement to the empirically calibrated dynamical model.}
\label{fig:causal_map}
\end{figure}

A key interpretive challenge lies in translating a nonlinear, data-driven
representation into a conceptually coherent account that remains faithful to
the empirical findings without overstating their scope. The discussion therefore
emphasises qualitative structure over coefficient magnitudes, and conditional
dependencies over universal causal claims. The stylised influence map presented
in Figure~\ref{fig:causal_map} reflects this approach: it highlights dominant and
interpretable pathways while omitting statistically detectable but conceptually
ambiguous links that would obscure the overall logic of the system.

\subsection{Overview of inferred causal structure}

The sparse regression results reveal a compact set of recurring dependencies among the ten system variables. Despite allowing for a high-dimensional nonlinear basis, only a limited number of terms persist after regularisation, and these terms cluster into a small number of coherent interaction patterns. This parsimony suggests that the observed dynamics are governed by a few dominant channels of influence rather than by diffuse cross-coupling, a finding that is consistent with recent arguments in the multi-system transitions literature emphasising structured interdependencies and patterned co-evolution over undifferentiated systemic complexity \citep{andersen2024multi,truffer2022perspective,donges2018taxonomies}.

The inferred causal structure can be understood as comprising three interacting domains. First, a social–technological domain centres on the co-evolution of clean innovation (V5) and social wellbeing (V8), which reinforce each other through mutually positive feedbacks. Second, an environmental and productive domain is organised around emissions (V2) and environmental stress (V9), with stress acting as a constraining influence on both renewable energy deployment (V1) and resource productivity (V3). Third, a governance and transition domain involves policy stringency (V6), transition finance (V7), and system-level resilience (V10). Within this domain, policy stringency supports financial mobilisation, finance enables innovation dynamics, and resilience conditions the system response by buffering environmental stress and stabilising social wellbeing, thereby shaping the context in which transformation unfolds.

Crucially, the use of flexible basis functions implies that higher-order terms cannot be dismissed a priori as negligible. Rather than representing marginal refinements, these terms often capture changes in system behaviour across regimes or the presence of saturation and constraint effects. Interpretation therefore focuses on the functional role of inferred dependencies, such as whether they act to reinforce, constrain, or condition system dynamics, rather than on the formal order of the underlying basis terms.

Within this framework, the diagram should be read as a compact summary of the dominant dependencies consistently identified by the sparse regression, not as an exhaustive causal map. Only relationships that are both empirically persistent and interpretable at the system level are shown. This selectivity reflects the purpose of the diagram: to clarify the main structural channels shaping the observed dynamics rather than to enumerate all possible interactions.

\subsection{Subsystems and feedbacks}

Several subsystems emerge as particularly salient in organising the system’s behaviour.

\paragraph{Innovation and wellbeing}
The most prominent reinforcing structure is the bidirectional relationship between clean innovation (V5) and social wellbeing (V8). Innovation contributes positively to wellbeing through improvements in environmental quality, service
provision, and economic opportunity, while higher levels of wellbeing stabilise the social, institutional, and cognitive conditions necessary for innovation, including skills, trust, and acceptance of change \citep{kohler2019agenda,truffer2022perspective,plank2021doing}. This feedback constitutes the clearest example of a mutually reinforcing dynamic in the system, consistent with transition scholarship emphasising co-evolutionary processes and reinforcing
mechanisms operating across technological and societal domains \citep{rosenbloom2017pathways,andersen2023multi}.

This result aligns closely with provisioning systems approaches and with socio-technical transition research, which emphasise that technological change is inseparable from social conditions. The inferred dynamics indicate that this reinforcement is nonlinear and subject to saturation: gains in wellbeing associated with innovation depend on the presence of complementary institutional capacity and material support, rather than arising automatically from innovation activity alone.

\paragraph{Environmental stress and productive capacity}
Environmental stress (V9) occupies a central role as a constraining variable. It increases with emissions (V2) and feeds back negatively onto both renewable energy uptake (V1) and resource productivity (V3). In this configuration, stress is not
merely an outcome variable but an active mediator that erodes system capacity and slows transformation.

Such behaviour is consistent with the literature on destabilising feedbacks in sustainability transitions, where environmental pressures undermine adaptive capacity, increase uncertainty, and divert resources toward short-term coping strategies \citep{scoones2020transformations,stirling2023transforming,turnheim2015evaluating}. The negative influence of stress on productivity and renewables suggests that sustained high-stress regimes can lock systems into reactive modes, limiting the effectiveness of technological and policy interventions
\citep{andersen2023multi,schluter2023unraveling}.

\paragraph{Transition finance and governance}
Transition finance (V7) plays a structurally important but conditional role within the system. In the inferred structure, finance primarily supports clean innovation (V5) and can be interpreted as one mechanism through which policy intent (V6) is translated into investment capacity. Its contribution to
wellbeing (V8) is therefore indirect and operates mainly through innovation dynamics, rather than appearing as a dominant stand-alone driver. This is consistent with transition research emphasising that finance becomes transformative when it is aligned with socio-technical change processes rather than treated as an exogenous enabling input \citep{kohler2019agenda,truffer2022perspective,andersen2023multi}.

This conditionality is reflected in the diagram, where the most salient pathway runs from policy to finance (V6 $\rightarrow$ V7) and from finance to innovation (V7 $\rightarrow$ V5), with downstream implications mediated via the innovation--wellbeing feedback (V5 $\leftrightarrow$ V8). More broadly, the
literature stresses that financial mobilisation can either accelerate or frustrate transitions depending on how it is institutionally structured and coupled to pathway governance \citep{rosenbloom2017pathways,turnheim2015evaluating,kanger2025integrated}.

Policy stringency (V6) enters as a steering variable that shapes finance, renewables, and emissions, but does not dominate the system’s dynamics. Its comparatively weak endogenous role likely reflects the absence of large, discrete policy shocks within the observed period and the coarse temporal resolution of the indicator, rather than an intrinsic lack of importance
\citep{van2016checklist,geels2022causality}.

\subsection{Interpretation in a multi-system transitions context}

Interpreted through a multi-system transformation (MST) lens, the inferred causal structure underscores the limits of linear or sector-specific accounts of sustainability transitions. The results suggest that transition pathways are shaped by the co-evolution of enabling and constraining forces across multiple domains, rather than by isolated technological substitution or single policy
levers \citep{kohler2019agenda,truffer2022perspective,andersen2023multi}.

The centrality of the innovation--wellbeing feedback highlights that social outcomes are not merely downstream consequences of transition, but integral components of its dynamics. Wellbeing both enables and is enabled by technological change, implying that social policy, labour conditions, and distributional effects are structurally relevant to transition performance rather than external constraints \citep{scoones2020transformations,plank2021doing}.

At the same time, the prominence of environmental stress as a constraining hub illustrates how cross-domain pressures can limit transformation even in the presence of innovation and finance. Stress links environmental conditions to productive capacity, reinforcing the argument that transitions cannot be understood or governed solely through emissions-focused metrics
\citep{donges2018taxonomies,schluter2023unraveling}.

Within this configuration, transition finance emerges as an endogenous, system-embedded variable whose influence is mediated through innovation and governance rather than operating as an autonomous driver. Financial conditions support or hinder transition trajectories primarily insofar as they enable
innovation, capital reallocation, and structural upgrading, consistent with pathway-oriented accounts of transition governance
\citep{rosenbloom2017pathways,turnheim2015evaluating,kanger2025integrated}. This interpretation avoids both over-attributing causal power to finance and treating it as a purely exogenous enabling input.

Overall, the inferred structure supports calls for integrated, cross-system modelling approaches that foreground feedbacks, constraints, and conditional dependencies, rather than relying on single-lever narratives or sector-isolated explanations of sustainability transitions
\citep{geels2022causality,schluter2024navigating}.

\subsection{Implications for policy and governance}

The diagnostic nature of the model suggests that policy implications should be formulated as structural steering principles rather than as optimisation recommendations \citep{kohler2018modelling,turnheim2015evaluating}.

First, the reinforcing relationship between innovation and wellbeing warrants particular attention. Policies that undermine social stability, increase inequality, or erode trust risk weakening this feedback and, by extension, slowing transition dynamics. Conversely, sustained investments in education,
health, and social infrastructure can indirectly accelerate innovation by maintaining favourable social and institutional conditions \citep{scoones2020transformations,plank2021doing}.

Second, environmental stress emerges as a variable that merits explicit governance attention. Stress reduction should not be treated solely as a by-product of emissions mitigation, but as a strategic objective in its own right. Policies that reduce emissions at the cost of heightened social or economic stress may ultimately prove counterproductive by eroding productive
and adaptive capacity \citep{donges2018taxonomies,schluter2023unraveling}.

Third, transition finance should be deployed as an enabling instrument rather than as a substitute for structural change. Financial flows are most effective when aligned with innovation and productivity improvements that also support social wellbeing. By contrast, large-scale investment in emissions-intensive or
purely compensatory infrastructure risks entrenching existing constraints without strengthening the social--technological core of the transition
\citep{rosenbloom2017pathways,kanger2025integrated}.

Finally, the comparatively indirect role of policy stringency cautions against overestimating direct control. Policy operates through the system rather than above it, and its effectiveness depends on timing, complementarities, and
alignment with social and technological conditions \citep{geels2022causality,truffer2022perspective}.

\subsection{Reflection and integration into the broader project context}

The sparse dynamical system presented here provides a quantitative foundation for a qualitative understanding of MST dynamics. By revealing a small number of persistent dependencies, it supports a modelling strategy focused on structural interdependencies rather than detailed forecasting.

In MST terms, the results correspond to a meso-level configuration of interacting subsystems that shape transition pathways. They justify an approach that combines quantitative diagnostics with scenario-based and participatory methods, using the
inferred causal structure as a guide rather than a prescription.

In summary, the findings reinforce a central insight of contemporary transition research: sustainability transitions hinge less on isolated technological change and more on the co-evolution of innovation, wellbeing, governance, and environmental constraints. This perspective aligns with the objectives of the
broader NEFI+ context, where socio-technical, institutional, and ecological dimensions must be evaluated jointly rather than sequentially.

\subsection{Scenario-dependent expression of the inferred dynamics}

A useful way to interpret the inferred signed dependencies and feedbacks is to treat
them as \emph{structural tendencies} whose realised effects depend on the institutional
and political configuration in which the transition unfolds. This aligns with the
transitions literature that frames futures as \emph{pathways} shaped by governance,
actor coordination, and sequencing, rather than as single trajectories determined by
technology cost curves or marginal incentives \citep{rosenbloom2017pathways, turnheim2015evaluating}.
It also aligns with multi-system perspectives that emphasise interdependence across
domains (energy, finance, social conditions, governance, environment) and the
possibility that acceleration in one domain can induce bottlenecks or rebound elsewhere
\citep{kohler2019agenda, andersen2023multi}.

In a \emph{market-led} scenario, the configuration is characterised by comparatively
weak directionality in governance and a reliance on decentralised allocation and
competition to deliver change. Transitions modelling reviews emphasise that such
configurations may achieve incremental progress, but often struggle with coordination
failures, distributional tensions, and lock-in effects when system change requires
aligned shifts across multiple domains \citep{kohler2018modelling, holtz2015prospects}.
Within this context, the inferred model structure suggests that beneficial loops (such
as mutually reinforcing innovation and wellbeing) may exist but remain vulnerable to
destabilising couplings (such as stress undermining productive capacity). In other
words, market-led configurations tend to expose the system to stronger sensitivity to
exogenous or endogenous stressors, because the institutions that buffer stress and
maintain legitimacy are not guaranteed by design \citep{scoones2020transformations}.

In a \emph{policy-driven / reform} scenario, stronger steering reduces uncertainty and
provides more coherent directionality, but the overall configuration may remain within
the boundaries of existing socio-technical and political-economic structures. In the
pathways literature, this corresponds to governance that prioritises measurability and
incremental adjustment, often improving near-term performance without necessarily
reconfiguring deeper interdependencies \citep{turnheim2015evaluating, rosenbloom2017pathways}.
In terms of the inferred dynamics, such a configuration would be expected to dampen
the more volatile feedbacks (e.g., those linked to stress) but may not fully leverage
the positive feedbacks unless policies explicitly align finance, innovation, and social
outcomes. This is consistent with broader agenda-setting contributions that caution
against over-reliance on single levers and call for explicit cross-domain coordination
\citep{kohler2019agenda, truffer2022perspective}.

In an \emph{authoritarian / crisis-response} scenario, coordination is achieved through
centralised authority and emergency logics. Transformations scholarship highlights that
crisis governance can reconfigure institutions rapidly, but may also weaken participation,
legitimacy, and adaptive capacity, especially when social outcomes are subordinated to
stabilisation objectives \citep{scoones2020transformations, stirling2023transforming}.
If environmental stress is treated as an external disturbance rather than a variable to
be governed, the inferred structure suggests a risk of persistent suppression of
productive and transition capacity, even if short-term volatility is reduced.

Finally, a \emph{collaborative / coordinated} scenario is defined by stronger alignment
across actors and domains, and by explicit attention to feedbacks between social
outcomes and technological change. Multi-system transitions research suggests that such
alignment can speed up net-zero transitions by activating reinforcing processes across
technologies, institutions, and actors \citep{andersen2023multi}. In terms of the model
structure, collaboration is best interpreted as a configuration that protects the
beneficial innovation--wellbeing loop while reducing the probability that stress
feedbacks dominate the system’s medium-run evolution. Across scenarios, the core point
is that differences are best understood as differences in \emph{feedback regimes} rather
than differences in single drivers, which is precisely why scenario work remains
complementary to empirical dynamical inference \citep{kohler2018modelling, jahel2023future}.

\subsection{Application and extension to Austria}

Although the present prototype is constructed at an aggregate European level, the approach is designed to be portable to national contexts, including Austria, provided the variable definitions and measurement choices are translated consistently. This is in line with methodological arguments that transition modelling should prioritise transparency, transferability, and legitimacy for policy use, rather than maximising structural detail at the expense of interpretability \citep{kohler2018modelling, van2016checklist}.
From a practical standpoint, Austria is an especially feasible candidate for extension because the majority of indicators used in this study are available in a suitable form at member state level from the same statistical sources used here.

At the level of inputs, Eurostat provides country-level series for renewable energy shares, resource productivity, digitalisation proxies, and EU-SILC based social indicators, which can be extracted for Austria with consistent definitions
\citep{V1_Eurostat_REShare, V3_Eurostat_ResourceProductivity, V4_Eurostat_Digitalisation, V8_Eurostat_Wellbeing}.
Similarly, emissions from ETS installations can be traced via the EEA viewer and filtered to Austrian installations or national aggregates \citep{V2_EEA_ETS_Emissions}. For governance and innovation proxies, OECD data infrastructure supports the use of
environmental policy stringency (EPS) and ENV-TECH patent indicators, which can be configured to national levels depending on coverage \citep{V6_OECD_PolicyStringency, V5_OECD_GreenPatents}.
Environmental stress and resilience proxies can likewise be adapted using Eurostat stress indicators and the JRC resilience dashboards, noting that national aggregation choices need to be documented carefully \citep{V9_Eurostat_EnvironmentalStress, V10_JRC_Resilience}.

Preliminary inspection of Austrian time series suggests that the country performs at or above the European average on several of the variables that are central to the reinforcing structures identified in the model. In particular, relatively high shares of renewable energy, stable social wellbeing indicators, and sustained innovation
activity point toward a configuration in which the innovation--wellbeing feedback is already comparatively well supported. These observations are indicative rather than diagnostic, but they suggest that Austria represents a context in which desirable
feedbacks are plausibly present rather than structurally constrained.

Conceptually, the main value of an Austrian application is therefore not to produce a national forecast, but to test whether the inferred \emph{dominant dependencies} remain stable when the system boundary is tightened. Multi-system transitions research
suggests that national contexts can differ in the strength and sign of inter-system couplings because institutional structures, actor configurations, and sectoral compositions vary \citep{andersen2023multi, lohr2023multi}. A national calibration
provides a structured way to examine whether the same feedback regimes appear, whether they are amplified or dampened, or whether additional couplings emerge through national governance arrangements or sectoral specialisation.

Methodologically, applying the model to Austria would also allow more deliberate integration with participatory and stakeholder-facing modelling processes. The literature on participatory modelling for sustainability pathways emphasises that
co-creation can improve salience and legitimacy, but requires careful structuring of assumptions and transparent handling of uncertainty \citep{van2019modeling,
moallemi2021evaluating}. A feasible extension would therefore combine the data-driven dynamical inference with stakeholder validation of variable meaning, scenario narratives, and the plausibility of key couplings, consistent with calls for mixed
quantitative--qualitative futures work in social--ecological systems
\citep{jahel2023future}. In practical terms, the Austrian case could serve as a pilot for a nested approach: an empirically constrained national backbone model, complemented by scenario workshops that interrogate how governance and finance
configurations might activate different feedback regimes.

\section{Conclusion and future work}

This feasibility study demonstrates how a compact, interpretable dynamical model can be used to extract dominant dependencies and feedback structures from macro-level transition indicators. The objective is not prediction, but the identification of
structural constraints and reinforcing or destabilising couplings that shape feasible pathways. This positioning is consistent with transition-modelling assessments that emphasise exploration, interpretability, and reflexivity, particularly where
uncertainty is deep and governance questions are central \citep{kohler2018modelling,
holtz2015prospects, truffer2022perspective}.

Three conclusions follow. First, the model-based feedback structure reinforces the view that sustainability transitions are inherently multi-system processes in which technological change, governance, finance, social conditions, and environmental
pressures co-evolve \citep{kohler2019agenda, andersen2023multi}. Second, the approach illustrates the practical value of combining empirical inference with explicit causal reasoning, while maintaining disciplinary humility about what causal claims can be
supported from observational time series \citep{geels2022causality, schluter2024navigating,
schluter2023unraveling}. Third, the results underscore that causal-loop representations are best treated as \emph{heuristic scaffolding} for scenario and governance dialogue,
not as definitive mechanism maps—an orientation aligned with recent critiques of overconfidence and ``model closure'' in policy-facing modelling \citep{stirling2023mind,
stirling2023transforming, foulds2025aligned}.

Future work should proceed along three complementary directions. One direction is \emph{methodological}: improving robustness checks and transparency around variable normalisation, basis selection, and sparsity choices, and benchmarking against other
transition-modelling approaches. Reviews of energy-transition modelling tools note that model diversity is unavoidable, and that credibility often depends on systematic comparison rather than on any single model form \citep{chang2021trends, susser2024rethink}.
A second direction is \emph{integration}: linking data-driven dynamical inference to participatory multi-modelling procedures that explicitly combine stakeholder knowledge,
scenario narratives, and cross-domain coupling hypotheses \citep{moallemi2021evaluating,
nespeca2024towards, van2019modeling}. A third direction is \emph{equity and legitimacy}: ensuring that model extensions explicitly address distributional consequences and the
politics of model use, consistent with recent sustainability science discussions on equity in modelling processes \citep{giang2024equity}.

Finally, an immediate applied step is to develop national pilots (e.g. Austria) based on adequate European data sources, documenting indicator choices and validating causal interpretations through stakeholder engagement \citep{van2016checklist, moallemi2021evaluating,
V1_Eurostat_REShare, V10_JRC_Resilience}. In summary, the present work contributes a pragmatic modelling template: structurally explicit, empirically anchored, and usable
as a backbone for scenario-based foresight and governance-oriented dialogue in multi-system transitions \citep{turnheim2015evaluating, rosenbloom2017pathways}.

\section*{Author contributions}
The study was jointly conceptualized by Sabin Roman, Klaus Kubeczko, and Vitaliy Soloviy through iterative discussions that shaped the research questions and overall analytical framework. Sabin Roman led the data curation, formal analysis, software development, visualization, and the preparation of the original manuscript draft, integrating methodological and empirical components of the study. Investigation and methodological development were conducted collaboratively by all authors, with continuous exchange throughout the research process. Klaus Kubeczko and Vitaliy Soloviy contributed substantially to funding acquisition, project administration, supervision, and the provision of resources, and provided sustained strategic and intellectual guidance. Writing of the manuscript was a collaborative process: Sabin Roman prepared the initial draft, while Klaus Kubeczko and Vitaliy Soloviy contributed through critical review, discussion, and editing. All authors engaged in sustained collective reflection on the findings, critically reviewed the draft, and approved the manuscript at this stage of the work.

For the purposes of later journal submission, author contributions are also reported using the CRediT taxonomy: Sabin Roman: Conceptualization; Data curation; Formal analysis; Investigation; Methodology; Software; Validation; Visualization; Writing – original draft. Klaus Kubeczko and Vitaliy Soloviy: Conceptualization; Investigation; Methodology; Validation; Funding acquisition; Project administration; Resources; Supervision; Writing – review \& editing.

\section*{Conflict of Interest Statement}
The authors declare no conflicts of interest.

\section*{Disclaimer}
\addcontentsline{toc}{section}{Disclaimer}
Co-funded by the European Union. Views and opinions expressed are however those of the author(s) only and do not necessarily reflect those of the European Union or European Research Executive Agency. Neither the European Union nor the granting authority can be held responsible for them.

\section*{Acknowledgements}
\addcontentsline{toc}{section}{Acknowledgements}
The authors gratefully acknowledge the AIT Austrian Institute of Technology as the main institutional supporter of this publication. AIT provided the main research environment, support, and facilitation for the work presented here. The authors also thank the STRN Methods School (Sustainability Transitions Research Network) for valuable methodological training and intellectual exchange that informed this work.

This publication is supported by the European Union's Horizon Europe research and innovation programme under the Marie Sk\l{}odowska-Curie Postdoctoral Fellowship Programme, SMASH co-funded under the grant agreement No.~101081355. The operation (SMASH project) is co-funded by the Republic of Slovenia and the European Union from the European Regional Development Fund.

\appendix
\section{Reproducible analysis code:\\ Sparse Orthogonal Regression Technique}
\label{app:code}


\paragraph{Code availability.}
The Python code and synthetic example data accompanying this appendix are available on Zenodo at DOI: \url{https://doi.org/10.5281/zenodo.20701582}. The repository provides an implementation of the Sparse Orthogonal Regression Technique (SORT) used for exploratory dynamical systems reconstruction from time-series data.

This appendix documents the Python implementation used to infer a sparse dynamical system from time-series data and to generate diagnostic plots comparing the inferred dynamics with a linear baseline. The script implements a sparse model discovery pipeline: it constructs a tensor-product library of candidate basis terms from the observed variables, fits sparse regression models to approximate the governing equations, and then simulates the resulting learned system forward in time. The included example data are synthetic but representative of the real-world indicator datasets used in the accompanying preprint: yearly-resolution time series with growth, saturation, and oscillatory dynamics spanning diverse industrial, energy, environmental, financial, technological, governance, and social dimensions.

\paragraph{Input and preprocessing.}
The script expects a comma-separated input file, specified by \texttt{CSV_PATH}, with a header row. The first column is time (\texttt{t_raw}), and the remaining columns are variables (\texttt{X_raw}). The variables are scaled to $[0,1]$ using a Min--Max transformation (\texttt{MinMaxScaler}). This ensures that basis evaluation and sparse regression are not dominated by differing physical units or magnitudes. The script then estimates numerical time derivatives $\dot{X}$ using finite differences (\texttt{compute_derivatives}), with central differences in the interior and one-sided differences at the endpoints.

\paragraph{Basis construction.}
To represent nonlinear interactions, the script builds a tensor-product basis (\texttt{build_tensor_basis}) from one-dimensional basis functions evaluated on each state component. Multi-indices up to total degree $d_{\max}$ are generated via \texttt{multi_indices_exact} and \texttt{multi_indices_upto}. The one-dimensional basis is selected through \texttt{BASIS_FUNC} and evaluated using \texttt{eval_1d_basis}. Supported basis families include Legendre polynomials, Laguerre polynomials, Laguerre functions, Hermite polynomials, rational Legendre functions, orthonormal rational Legendre functions, and a Fourier basis using sine and cosine terms. The tensor-product library contains all products of one-dimensional basis terms whose total degree does not exceed $d_{\max}$, yielding an interpretable set of candidate interactions among variables.

\paragraph{Sparse regression and model identification.}
For each variable $x_j$, the code fits a sparse model of the form
$\dot{x}_j \approx \Phi(x)\beta_j + b_j$, where $\Phi(x)$ is the candidate basis matrix and $\beta_j$ is a sparse coefficient vector. Sparsity is enforced using LASSO regression (\texttt{sklearn.linear_model.Lasso}) in \texttt{fit_sparse_models}, using a time-ordered train/test split (\texttt{shuffle=False}) to respect temporal structure. The script reports the number of nonzero coefficients and the train/test RMSE for each equation. The resulting set of sparse models defines the learned vector field of the dynamical system.

\paragraph{Dynamical-system evaluation and forward simulation.}
The learned vector field is evaluated by \texttt{learned_rhs}, which reconstructs $\Phi(x)$ at a given state $x$ and computes each $\dot{x}_j$ from the corresponding sparse model. Forward simulation uses a standard explicit Runge--Kutta method: \texttt{rk4_step} performs a single fourth-order Runge--Kutta step and \texttt{simulate_system} iterates over an extended time grid. Time is extended by a user-defined horizon (\texttt{EXTRA_YEARS}) using the mean observed time step.

\paragraph{Back-transformation and baseline comparison.}
Simulated trajectories are transformed back to raw units using the inverse of the Min--Max scaler. As a simple baseline, the script fits independent linear regressions $x_j(t) \approx a_j t + b_j$ for each variable (\texttt{linear_regression_forecast}). The functions \texttt{plot_normalized} and \texttt{plot_model_vs_linear} generate diagnostic plots showing observed data, learned dynamical-system trajectories, forward simulations, and linear baselines. Training and test points are shown in different colours, and figures can be automatically saved as PNG files for selected variables.

\paragraph{Sparse term reporting.}
The repository also includes a sparse term contribution reporter. This utility lists the nonzero basis terms selected for each learned equation and formats them according to the selected basis family rather than as ordinary monomials. For example, terms are displayed as Legendre, rational Legendre, Laguerre, Hermite, or Fourier basis functions depending on \texttt{BASIS_FUNC}. This supports inspection of the learned equations and helps make the sparse dynamical structure more transparent.

\bibliographystyle{elsarticle-harv} 
\bibliography{biblio}

@article{roman2026utci,
  author  = {Roman, Sabin and Todorovski, Ljup{\v{c}}o and D{\v{z}}eroski, Sa{\v{s}}o and Skok, Gregor},
  title   = {Approximating the universal thermal climate index using sparse regression with orthogonal polynomials},
  journal = {Geoscientific Model Development},
  volume  = {19},
  pages   = {4319--4330},
  year    = {2026},
  doi     = {10.5194/gmd-19-4319-2026}
}

@misc{feng2026illconditioning,
  author        = {Feng, Yuxiang and Mangan, Niall M. and Jayadharan, Manu},
  title         = {Ill-Conditioning in Dictionary-Based Dynamic-Equation Learning: A Systems Biology Case Study},
  year          = {2026},
  eprint        = {2603.11330},
  archivePrefix = {arXiv},
  primaryClass  = {q-bio.QM},
  doi           = {10.48550/arXiv.2603.11330}
}

@article{kohler2018modelling,
  title={Modelling sustainability transitions: An assessment of approaches and challenges},
  author={K{\"o}hler, Jonathan and de Haan, Fjalar and Holtz, Georg and Kubeczko, Klaus and Moallemi, Enayat and Papachristos, George and Chappin, Emile},
  journal={Journal of Artificial Societies and Social Simulation},
  volume={21},
  number={2},
  pages={1--20},
  year={2018},
  doi={10.18564/jasss.3629}
}

@article{kohler2019agenda,
  title={An agenda for sustainability transitions research: State of the art and future directions},
  author={K{\"o}hler, Jonathan and Geels, Frank W and Kern, Florian and Markard, Jochen and Onsongo, Elsie and Wieczorek, Anna and Alkemade, Floortje and Avelino, Flor and Bergek, Anna and Boons, Frank and others},
  journal={Environmental Innovation and Societal Transitions},
  volume={31},
  pages={1--32},
  year={2019},
  publisher={Elsevier},
  doi={10.1016/j.eist.2019.01.004}
}

@article{truffer2022perspective,
  title={A perspective on the future of sustainability transitions research},
  author={Truffer, Bernhard and Rohracher, Harald and Kivimaa, Paula and Raven, Rob and Alkemade, Floor and Carvalho, Luis and Feola, Giuseppe},
  journal={Environmental Innovation and Societal Transitions},
  volume={42},
  pages={331--339},
  year={2022},
  publisher={Elsevier}
}

@article{rosenbloom2017pathways,
  title={Pathways: An emerging concept for the theory and governance of low-carbon transitions},
  author={Rosenbloom, Daniel},
  journal={Global Environmental Change},
  volume={43},
  pages={37--50},
  year={2017},
  publisher={Elsevier}
}

@article{turnheim2015evaluating,
  title={Evaluating sustainability transitions pathways: Bridging analytical approaches to address governance challenges},
  author={Turnheim, Bruno and Berkhout, Frans and Geels, Frank and Hof, Andries and McMeekin, Andy and Nykvist, Bj{\"o}rn and van Vuuren, Detlef},
  journal={Global Environmental Change},
  volume={35},
  pages={239--253},
  year={2015},
  publisher={Elsevier}
}

@incollection{stirling2023mind,
  title={Mind the unknowns: Exploring the politics of ignorance in mathematical models},
  author={Stirling, Andy},
  booktitle={The Politics of Modelling: Numbers Between Science and Policy},
  editor={Saltelli, Andrea and Di Fiore, Monica},
  pages={99--118},
  year={2023},
  publisher={Oxford University Press},
  doi={10.1093/oso/9780198872412.003.0007}
}

@article{jahel2023future,
  title={The future of social-ecological systems at the crossroads of quantitative and qualitative methods},
  author={Jahel, Camille and Bourgeois, Robin and Bourgoin, Jeremy and De Lattre-Gasquet, Marie and Delay, Etienne and Dumas, Patrice and Le Page, Christophe and Piraux, Marc and Prudhomme, R{\'e}mi and others},
  journal={Technological Forecasting and Social Change},
  volume={193},
  pages={122624},
  year={2023},
  publisher={Elsevier}
}

@article{plank2021doing,
  title={Doing more with less: Provisioning systems and the transformation of the stock-flow-service nexus},
  author={Plank, Christina and Liehr, Stefan and Hummel, Diana and Wiedenhofer, Dominik and Haberl, Helmut and G{\"o}rg, Christoph},
  journal={Ecological Economics},
  volume={187},
  pages={107093},
  year={2021},
  publisher={Elsevier}
}

@techreport{andersen2024multi,
  title={Multi-system dynamics in sustainability transitions: Introduction and outlook},
  author={Andersen, Allan Dahl and Markard, Jochen},
  institution={Cambridge University Press},
  series={Cambridge Open Engage},
  type={Working paper},
  address={United Kingdom},
  year={2024},
  doi={10.33774/coe-2024-x6x8n}
}

@article{foulds2025aligned,
  title={Aligned interpretations? Comparing energy modeller and policymaker perspectives on model development and use},
  author={Foulds, Chris and Jones, Aled and Royston, Sarah and Pasqualino, Roberto},
  journal={Energy Reports},
  volume={14},
  pages={1866--1876},
  year={2025},
  publisher={Elsevier}
}

@article{stirling2023transforming,
  title={Transforming imaginations? Multiple dimensionalities and temporalities as vital complexities in transformations to sustainability},
  author={Stirling, Andy and Cairns, Rose and Johnstone, Phil and Onyango, Joel},
  journal={Global Environmental Change},
  volume={82},
  pages={102741},
  year={2023},
  publisher={Elsevier}
}

@article{scoones2020transformations,
  title={Transformations to sustainability: combining structural, systemic and enabling approaches},
  author={Scoones, Ian and Stirling, Andrew and Abrol, Dinesh and Atela, Joanes and Charli-Joseph, Lakshmi and Eakin, Hallie and Ely, Adrian and Olsson, Per and Pereira, Laura and Priya, Ritu and others},
  journal={Current Opinion in Environmental Sustainability},
  volume={42},
  pages={65--75},
  year={2020},
  publisher={Elsevier}
}

@article{kanger2025integrated,
  title={Integrated framework of intervention points and transformative outcomes for single-and multi-system transitions},
  author={Kanger, Laur and Ghosh, Bipashyee and Entsalo, Hanna},
  journal={Technological Forecasting and Social Change},
  volume={216},
  pages={124146},
  year={2025},
  publisher={Elsevier},
  doi={10.1016/j.techfore.2025.124146}
}

@article{moallemi2020exploratory,
  title={Exploratory modeling for analyzing coupled human-natural systems under uncertainty},
  author={Moallemi, Enayat A and Kwakkel, Jan and de Haan, Fjalar J and Bryan, Brett A},
  journal={Global Environmental Change},
  volume={65},
  pages={102186},
  year={2020},
  publisher={Elsevier}
}

@article{chang2021trends,
  title={Trends in tools and approaches for modelling the energy transition},
  author={Chang, Miguel and Thellufsen, Jakob Zink and Zakeri, Behnam and Pickering, Bryn and Pfenninger, Stefan and Lund, Henrik and {\O}stergaard, Poul Alberg},
  journal={Applied Energy},
  volume={290},
  pages={116731},
  year={2021},
  publisher={Elsevier}
}

@article{donges2018taxonomies,
  title={Taxonomies for structuring models for {World--Earth} systems analysis of the {Anthropocene}: subsystems, their interactions and social--ecological feedback loops},
  author={Donges, Jonathan F. and Lucht, Wolfgang and Cornell, Sarah E. and Heitzig, Jobst and Barfuss, Wolfram and Lade, Steven J. and Schl{\"u}ter, Maja},
  journal={Earth System Dynamics},
  volume={12},
  number={4},
  pages={1115--1137},
  year={2021},
  doi={10.5194/esd-12-1115-2021}
}

@article{schluter2024navigating,
  title={Navigating causal reasoning in sustainability science},
  author={Schl{\"u}ter, Maja and Hertz, Tilman and Mancilla Garc{\'\i}a, Mar{\'\i}a and Banitz, Thomas and Grimm, Volker and Johansson, Lars-G{\"o}ran and Lindkvist, Emilie and Mart{\'\i}nez-Pe{\~n}a, Rodrigo and Radosavljevic, Sonja and Wennberg, Karl and others},
  journal={Ambio},
  volume={53},
  number={11},
  pages={1618--1631},
  year={2024},
  publisher={Springer}
}

@article{geels2022causality,
  title={Causality and explanation in socio-technical transitions research: Mobilising epistemological insights from the wider social sciences},
  author={Geels, Frank W},
  journal={Research Policy},
  volume={51},
  number={6},
  pages={104537},
  year={2022},
  publisher={Elsevier},
  doi={10.1016/j.respol.2022.104537}
}

@article{schluter2023unraveling,
  title={Unraveling complex causal processes that affect sustainability requires more integration between empirical and modeling approaches},
  author={Schl{\"u}ter, Maja and Brelsford, Christa and Ferraro, Paul J and Orach, Kirill and Qiu, Minghao and Smith, Martin D},
  journal={Proceedings of the National Academy of Sciences},
  volume={120},
  number={41},
  pages={e2215676120},
  year={2023},
  publisher={National Academy of Sciences},
  doi={10.1073/pnas.2215676120}
}

@article{van2016checklist,
  title={A checklist for model credibility, salience, and legitimacy to improve information transfer in environmental policy assessments},
  author={van Voorn, George AK and Verburg, Ren{\'e} W and Kunseler, E-M and Vader, Janneke and Janssen, Peter HM},
  journal={Environmental Modelling \& Software},
  volume={83},
  pages={224--236},
  year={2016},
  publisher={Elsevier}
}

@article{andersen2023multi,
  title={Multi-system dynamics and the speed of net-zero transitions: Identifying causal processes related to technologies, actors, and institutions},
  author={Andersen, Allan Dahl and Geels, Frank W},
  journal={Energy Research \& Social Science},
  volume={102},
  pages={103178},
  year={2023},
  publisher={Elsevier}
}

@article{lohr2023multi,
  title={Multi-system interactions in hydrogen-based sector coupling projects: System entanglers as key actors},
  author={L{\"o}hr, Meike and Chlebna, Camilla},
  journal={Energy Research \& Social Science},
  volume={105},
  pages={103282},
  year={2023},
  publisher={Elsevier},
  doi={10.1016/j.erss.2023.103282}
}

@article{allouche2024nexus,
  title={Nexus framing of sustainability issues: feasibility, synergies, and trade-offs in terms of water-energy-food},
  author={Allouche, Jeremy},
  journal={Annual Review of Environment and Resources},
  volume={49},
  number={1},
  pages={501--518},
  year={2024},
  publisher={Annual Reviews}
}

@article{nespeca2024towards,
  title={Towards participatory multi-modeling for policy support across domains and scales: a systematic procedure for integral multi-model design},
  author={Nespeca, Vittorio and Quax, Rick and Rikkert, Marcel GM and Korzilius, Hubert PLM and Marchau, Vincent AWJ and Hadijsotiriou, Sophie and Oreel, Tom and Coenen, Jannie and Wertheim, Heiman and Voinov, Alexey and others},
  journal={arXiv preprint arXiv:2402.06228},
  year={2024}
}

@article{van2019modeling,
  title={Modeling with stakeholders for transformative change},
  author={van Bruggen, Anne and Nikolic, Igor and Kwakkel, Jan},
  journal={Sustainability},
  volume={11},
  number={3},
  pages={825},
  year={2019},
  publisher={MDPI}
}

@article{moallemi2021evaluating,
  title={Evaluating participatory modeling methods for co-creating pathways to sustainability},
  author={Moallemi, Enayat A and de Haan, Fjalar J and Hadjikakou, Michalis and Khatami, Sina and Malekpour, Shirin and Smajgl, Alex and Smith, M Stafford and Voinov, Alexey and Bandari, Reihaneh and Lamichhane, Prahlad and others},
  journal={Earth's Future},
  volume={9},
  number={3},
  pages={e2020EF001843},
  year={2021},
  publisher={Wiley Online Library}
}

@article{giang2024equity,
  title={Equity and modeling in sustainability science: Examples and opportunities throughout the process},
  author={Giang, Amanda and Edwards, Morgan R and Fletcher, Sarah M and Gardner-Frolick, Rivkah and Gryba, Rowenna and Mathias, Jean-Denis and Venier-Cambron, Camille and Anderies, John M and Berglund, Emily and Carley, Sanya and others},
  journal={Proceedings of the National Academy of Sciences},
  volume={121},
  number={13},
  pages={e2215688121},
  year={2024},
  publisher={National Academy of Sciences},
  doi={10.1073/pnas.2215688121}
}

@incollection{susser2024rethink,
  title={Rethink energy system models to support interdisciplinary and inclusive just transition debates},
  author={S{\"u}sser, Diana and McGookin, Connor and McDowall, Will and Lombardi, Francesco and Braunreiter, Lukas and Bouzarovski, Stefan},
  booktitle={Strengthening European Energy Policy},
  editor={Crowther, Ami and Foulds, Chris and Robison, Rosie and Gladkykh, Ganna},
  pages={145--157},
  year={2024},
  publisher={Palgrave Macmillan},
  address={Cham},
  doi={10.1007/978-3-031-66481-6_11}
}

@article{papachristos2011system,
  title={A system dynamics model of socio-technical regime transitions},
  author={Papachristos, Georg},
  journal={Environmental Innovation and Societal Transitions},
  volume={1},
  number={2},
  pages={202--233},
  year={2011},
  publisher={Elsevier}
}

@article{holtz2015prospects,
  title={Prospects of modelling societal transitions: Position paper of an emerging community},
  author={Holtz, Georg and Alkemade, Floortje and De Haan, Fjalar and K{\"o}hler, Jonathan and Trutnevyte, Evelina and Luthe, Tobias and Halbe, Johannes and Papachristos, George and Chappin, Emile and Kwakkel, Jan and others},
  journal={Environmental Innovation and Societal Transitions},
  volume={17},
  pages={41--58},
  year={2015},
  publisher={Elsevier}
}

@article{knobloch2016behavioural,
  title={The behavioural aspect of green technology investments: A general positive model in the context of heterogeneous agents},
  author={Knobloch, Florian and Mercure, Jean-Francois},
  journal={Environmental Innovation and Societal Transitions},
  volume={21},
  pages={39--55},
  year={2016},
  publisher={Elsevier}
}

@misc{V1_Eurostat_REShare,
  author       = {{Eurostat}},
  title        = {Renewable Energy Share in Gross Final Energy Consumption (SHARES)},
  year         = {2024},
  note         = {Dataset nrg\_ind\_ren, European Union aggregate},
  howpublished = {\url{https://ec.europa.eu/eurostat/databrowser/view/nrg_ind_ren__custom_19010951/default/table}},
  institution  = {European Commission}
}

@misc{V2_EEA_ETS_Emissions,
  author       = {{European Environment Agency}},
  title        = {{EU} Emissions Trading System: Verified Emissions from Stationary Installations},
  year         = {2024},
  note         = {Power and industrial installations, EU aggregate},
  howpublished = {\url{https://www.eea.europa.eu/en/analysis/maps-and-charts/emissions-trading-viewer-1-dashboards}},
  institution  = {European Environment Agency}
}

@misc{V3_Eurostat_ResourceProductivity,
  author       = {{Eurostat}},
  title        = {Resource Productivity: {GDP} per Unit of Domestic Material Consumption},
  year         = {2024},
  note         = {Dataset env\_ac\_rp, EU aggregate},
  howpublished = {\url{https://ec.europa.eu/eurostat/databrowser/view/env_ac_rp/default/line}},
  institution  = {European Commission}
}

@misc{V4_Eurostat_Digitalisation,
  author       = {{Eurostat}},
  title        = {E-commerce Usage by Enterprises},
  year         = {2024},
  note         = {Share of enterprises with e-commerce sales, EU},
  howpublished = {\url{https://ec.europa.eu/eurostat/statistics-explained/index.php?title=E-commerce_statistics}},
  institution  = {European Commission}
}

@misc{V5_OECD_GreenPatents,
  author       = {{OECD}},
  title        = {Environment-related Technologies (ENV-TECH) Patent Indicators},
  year         = {2024},
  note         = {Share of environment-related patent families, EU aggregate},
  howpublished = {\url{https://data-explorer.oecd.org/}},
  institution  = {Organisation for Economic Co-operation and Development}
}

@misc{V6_OECD_PolicyStringency,
  author       = {{OECD}},
  title        = {Environmental Policy Stringency Index},
  year         = {2024},
  note         = {EPS index, EU aggregate},
  howpublished = {\url{https://data-explorer.oecd.org/}},
  institution  = {Organisation for Economic Co-operation and Development}
}

@misc{V7_EEA_GreenBonds,
  author       = {{European Environment Agency}},
  title        = {Green Bonds and Sustainable Finance Indicators},
  year         = {2024},
  note         = {Green bond issuance and sustainable finance flows, EU},
  howpublished = {\url{https://www.eea.europa.eu/en/analysis/indicators/green-bonds-8th-eap}},
  institution  = {European Environment Agency}
}

@misc{V8_Eurostat_Wellbeing,
  author       = {{Eurostat}},
  title        = {Life Satisfaction and Economic Strain Indicators (EU-SILC)},
  year         = {2024},
  note         = {Subjective well-being and material deprivation indicators},
  howpublished = {\url{https://ec.europa.eu/eurostat/databrowser/view/ILC_PW01/default/table}},
  institution  = {European Commission}
}

@misc{V9_Eurostat_EnvironmentalStress,
  author       = {{Eurostat}},
  title        = {Water Exploitation Index (WEI+) and Climate-related Environmental Stress Indicators},
  year         = {2024},
  note         = {SDG 6.4.2 and related indicators, EU aggregate},
  howpublished = {\url{https://ec.europa.eu/eurostat/databrowser/view/sdg_06_60/default/line}},
  institution  = {European Commission}
}

@misc{V10_JRC_Resilience,
  author       = {{European Commission Joint Research Centre}},
  title        = {Resilience Dashboards for Energy, Food, and Infrastructure Systems},
  year         = {2024},
  note         = {Composite resilience indicators for EU provisioning systems},
  howpublished = {\url{https://datam.jrc.ec.europa.eu/datam/mashup/RESILIENCE_DASHBOARDS/}},
  institution  = {European Commission}
}

@article{brunton2016discovering,
  title={Discovering governing equations from data by sparse identification of nonlinear dynamical systems},
  author={Brunton, Steven L. and Proctor, Joshua L. and Kutz, J. Nathan},
  journal={Proceedings of the National Academy of Sciences},
  volume={113},
  number={15},
  pages={3932--3937},
  year={2016},
  doi={10.1073/pnas.1517384113}
}

@article{galeotti2020measuring,
  title={Measuring environmental policy stringency: Approaches, validity, and impact on environmental innovation and energy efficiency},
  author={Galeotti, Marzio and Salini, Silvia and Verdolini, Elena},
  journal={Energy Policy},
  year={2020},
  volume={136},
  pages={111052},
  doi={10.1016/j.enpol.2019.111052}
}

@article{smolenska2025european,
  title={European capitalisms in sustainability transition: the case of green bonds},
  author={Smole{\'n}ska, Agnieszka},
  journal={Journal of European Public Policy},
  year={2025},
  doi={10.1080/13501763.2025.2521395}
}



\end{document}